\documentclass[final]{dmtcs-episciences}

\usepackage[utf8]{inputenc}
\usepackage{subfigure}

\usepackage{pdflscape}
\usepackage{geometry}

\usepackage[round]{natbib}

\usepackage{booktabs,xcolor,colortbl,multirow,rotating,bigstrut}
\usepackage{listings}
\usepackage{amsmath,amsfonts,amsthm,amssymb}
\usepackage[nameinlink]{cleveref}
\usepackage{adjustbox}
\newtheorem{conj}{Conjecture}

\usepackage{tikz}

\definecolor{mygreen}{rgb}{0,0.6,0}
\lstdefinelanguage{essence}{
 frame = single,
 breaklines=true,
 keywords = { language, relation, Essence, given, letting, find, such, that, function, total, surjective, be , forAll, exists, injective, in, preImage, range ,  mset, set, partition, new, type, intersect, from , minimising, maximising, indexed, by, defined, maxSize, maxNumParts , subset , size, toInt, sum , sequence },
  keywordstyle=\color{blue}\bfseries,
  ndkeywords={atleast, atmost, gcc, int, matrix, bool},
  ndkeywordstyle=\color{mygreen}\bfseries,
  identifierstyle=\color{black},
  sensitive=false,
  comment=[l]{\$},
  commentstyle=\color{violet}\rmfamily,
  stringstyle=\color{teal}\ttfamily,
  morestring=[b]',
  morestring=[b]",
}

\author[Ruth Hoffmann et al.]{Ruth Hoffmann\affiliationmark{1}
  \and \"{O}zg\"{u}r Akg\"{u}n\affiliationmark{1}
  \and Christopher Jefferson\affiliationmark{2,3}}
\title[Composable Constraint Models for Permutation Enumeration]{Composable Constraint Models for Permutation Enumeration}
\affiliation{
  School of Computer Science, University of St Andrews, St Andrews, Scotland \\
  School of Computer Science and Engineering, Central South University, Changsha, PR China\\
  Computing, School of Science \& Engineering, University of Dundee, Dundee, Scotland
  }
\keywords{Constraint Modelling, Permutation Pattern, Enumeration}
\begin{document}
\publicationdata{vol. 26:1, Permutation Patterns 2023}{2025}{6}{10.46298/dmtcs.12620}{2023-11-30; 2023-11-30; 2024-10-21; 2025-01-09}{2025-01-09}

\maketitle

\begin{abstract}

Constraint programming (CP) is a powerful tool for modeling mathematical concepts and objects and finding both solutions or counterexamples. One of the major strengths of CP is that problems can easily be combined or expanded. In this paper, we illustrate that this versatility makes CP an ideal tool for exploring problems in permutation patterns.

We declaratively define permutation properties, permutation pattern avoidance and containment constraints using CP and show how this allows us to solve a wide range of problems.
We show how this approach enables the arbitrary composition of these conditions, and also allows the easy addition of extra conditions.
We demonstrate the effectiveness of our techniques by modelling the containment and avoidance of six permutation patterns, eight permutation properties and measuring five statistics on the resulting permutations. 
In addition to calculating properties and statistics for the generated permutations, we show that arbitrary additional constraints can also be easily and efficiently added.

This approach enables mathematicians to investigate permutation pattern problems in a quick and efficient manner.
We demonstrate the utility of constraint programming for permutation patterns by showing how we can easily and efficiently extend the known permutation counts for a conjecture involving the class of $1324$ avoiding permutations.
For this problem, we expand the enumeration of $1324$-avoiding permutations with a fixed number of inversions to permutations of length 16 and show for the first time that in the enumeration there is a pattern occurring which follows a unique sequence on the Online Encyclopedia of Integer Sequences.
\end{abstract}

\section{Introduction}
The concept of permutation pattern classes emerged from an exercise question in the Art of Computer Programming Volume 1 (Section 2.2.1, Exercise 5) by \cite{knuth}, in which Knuth explored which permutations can be sorted in a limited stack-based system, called stack-sortable permutations.
 \cite{simion1985restricted} were among the pioneers in characterizing and enumerating sets of permutations based on the patterns that they avoid.
Further enumeration results garnered interest because the number of the original set of permutations, identified in Knuth's exercise, proved to be the Catalan numbers.

The enumeration results and subsequent research were primarily motivated by a conjecture proposed by Herbert Wilf at the 1992 SIAM meeting.
This conjecture stated that every permutation pattern class that avoids one permutation pattern exhibits an exponential growth rate. \cite{marcus2004excluded} later confirmed and proved this conjecture.

There are several systems which use computational techniques to assist with the enumeration of permutation classes and sets, some examples include Permuta \cite{Permuta}, PatternClass \cite{PatternClass}, Combinatorial Specification Searcher \cite{CombExp} and PermCode (previously PermLab) \cite{albert2012permlab}.
Permuta is a Python package which supports the enumeration of permutation classes using the BiSC algorithm by \cite{magnusson2012algorithms}.
PatternClass is a \cite{GAP4} package that encodes the permutations and classes into regular languages, enabling efficient investigation of permutation sets.
Combinatorial Specification Searcher is another Python package that provides a more exploratory approach by asking the user to define strategies on how to build combinatorial sets from other sets.
It combines these into a general enumeration algorithm.
PermCode is a Java library with a graphical user interface which allows for the exploration of the sets of permutations which avoid classical patterns.

Constraint programming (CP) is a generic and powerful paradigm that allows the expression of complex combinatorial problems in a declarative manner. 
In CP problems are expressed declaratively by giving a list of variables whose values must be found, and stating relationships between variables in the form of constraints. 
This declarative expression is called a model. 
CP solvers take this declarative version of the problem and use a range of different algorithms to solve the problem efficiently. 
The primary advantage of CP is that it allows the user to focus on the problem formulation, while the computer takes responsibility for the problem-solving methodology.

In the context of studying permutation patterns, constraint programming can offer significant benefits. 
Permutation patterns provide a prominent and widely researched topic within the domain of combinatorial mathematics. 
These patterns, which involve the arrangement of numbers in a certain order, exhibit intricate and complex structures. 
Constraint programming allows us to model these structures and constraints explicitly, enabling systematic and efficient exploration of the solution space.

In this paper we express problems in the high-level constraint language Essence by \cite{frisch2008ssence}. 
Essence, as implemented by the system Conjure by \cite{akgun2022conjure}, provides more abstract representation of combinatorial problems. 
Conjure converts high-level specifications given in Essence into a range of different formats, allowing many different constraint solvers (and related technologies, such as Mixed Integer Programming and Boolean Satisfiability) to be used when solving the written (also called \emph{modelled}) constraint problem.
This automation not only reduces the possibility of errors in translation but also frees the user to concentrate on the higher-level aspects of problem modelling. 
In this paper, we mainly use the constraint solver Minion created by \cite{minion}, which has been used to solve several large scale combinatorial problems in the past, including the number of semi-groups of order 10 \cite{semigroupcount} and for the enumeration of set-theoretic solutions to the Yang-Baxter equation \cite{akgun2022enumeration}.

In this paper we show how CP can aid the research in permutation patterns through the following contributions:
\begin{enumerate}
    \item A systematic and comprehensive treatment of 6 kinds of permutation pattern (as both avoidance and containment) as well as 13 properties and 5 statistics. Each one is implemented as a standalone constraint model. This is presented in \Cref{sec:lib}.
    \item The models of the patterns, properties and statistics are structured in a way that allows seamless composition of the individual models. \Cref{sec:composability} illustrates this compositionality.
    \item An evaluation of the flexible compositional constraint programming approach in 2 settings, one extended illustrative example and one example showing how composable CP models can extend current results such as a conjecture in \cite{CLAESSON20121680}. In \Cref{sec:figex} and \Cref{sec:avinv} we present the examples.
    \item 2 new conjectures resulting from the model enumerating $1324$-avoiding permutations with a fixed number of inversions which come from extending the computational results in \cite{CLAESSON20121680}.
\end{enumerate}

We believe our library of models, which can be easily composed in a "pick and mix" fashion, will further the research in permutation patterns by allowing for more efficient computational experimentation and exhaustive search as the constraint programming approach avoids the need for the traditional "generate-and-test" approach, which generates intermediate assignments that only satisfy some of the properties for later filtering.

This "generate-and-test" approach, which requires specialised algorithms to be  created to find all permutations which satisfy some given property, can work very well for that property but this approach is difficult to extend and requires considerable expertise in programming to add new properties. 
On the other hand, combining CP models can immediately allow highly efficient searching for a combination of properties, and our experience is adding new patterns in CP is significantly easier than create new bespoke state-of-the-art programs for searching for new patterns.

We choose not to compare our models against tools such as PermLab~\cite{albert2012permlab}, Combinatorial Specification Searcher~\cite{CombExp} or PatternClass~\cite{PatternClass}, as to our knowledge these tools do not support all features that we implement (for example number of inversions, parity, block-wise simplicity), or have a specialised focus (for example on regular classes) than the library of models which we will present.

\section{Introduction to Constraint Programming}

Constraint Programming (CP) is a powerful paradigm for solving complex combinatorial problems by focusing on the formulation of constraints rather than explicit solution algorithms. CP is particularly useful in situations where you need to find solutions that satisfy a set of conditions or optimise a certain objective under constraints.

The process of working in CP typically involves three main steps:
\begin{itemize}
    \item Modelling the problem: This involves defining the problem in a high-level language such as Essence \cite{frisch2008ssence}, where variables and constraints are described abstractly, without specifying how to solve the problem.
    \item Translation to a solver-friendly format: The high-level model is then automatically translated into a lower-level format that can be handled by a solver. Tools like Conjure \cite{akgun2022conjure} facilitate this step by converting the Essence model into formats compatible with solvers such as Minion \cite{minion}, Chuffed \cite{chu2018chuffed}, or MiniSAT \cite{een2003extensible}.
    \item Solving the problem: The translated model is processed by a solver, which systematically searches for a solution by trial and error, backtracking when constraints are violated, until a valid solution is found.
\end{itemize}

\begin{lstlisting}[ language=essence,caption=A tour of the most relevant features of the problem specification language Essence,label=lst:essence, float,float,floatplacement=t]
given <name> : <domain>   $ declaring a parameter
letting <name> be <value> $ setting the value of a parameter or a local alias
find <name> : <domain>    $ declaring a decision variable
such that <constraint>    $ posting a constraint 

$ specific examples from models used in this paper
given length : int
given avoidances : set of sequence of int
find perm : sequence (size length, injective) of int(1..length)

$ constraint types
such that forAll i : int(1..length) . <constraint>
such that forAll m : matrix indexed by [int(1..pattern)] of int(1..length) . <constraint>
such that forAll q : <domain> , <condition> . <constraint>
such that forAll av in avoidances . <constraint>
such that perm(1) = 2, perm(2) = 3
\end{lstlisting}

Most users of CP interact primarily with the modelling step, where they define the problem abstractly in a language like Essence. The remaining steps—translation and solving—are typically handled by automated tools.

\subsection{Constraint Modelling}

Constraint modelling in CP consists of representing the problem using a set of variables, parameters, and constraints, all described declaratively. This allows the solver to explore the search space and identify solutions that satisfy all of the specified constraints. In \Cref{lst:essence} is a sample Essence code that demonstrates key constructs and patterns in constraint modelling:

We first start by an explanation of key constructs in Essence, using the example fragments in \Cref{lst:essence}.

Declaring parameters (line 1): The \textit{given} keyword defines parameters that act as fixed inputs to the model. Parameters have a name and a domain. For instance, on line 6, length is declared as an integer parameter that will specify the length of the permutation. Parameters allow for flexibility, as they can be modified without changing the model's core structure.

Setting values (line 2): The \textit{letting} keyword assigns a value to a parameter or local alias. This can be used for simplification or to define constants. In practice, this makes it easier to reference certain values throughout the model.

Declaring decision variables (line 3): The \textit{find} keyword is used to define decision variables, which are the unknowns that the solver will try to assign values to. On line 8, for example, the variable perm is declared as a sequence, and it is required to be injective (meaning all elements in the sequence must be unique).

Adding constraints (Line 4): The \textit{such that} keyword introduces constraints, which are conditions that must be satisfied by the decision variables. These constraints guide the solver to ensure valid solutions. On line 14, a constraint explicitly fixes certain values of the permutation, forcing the first element to be 2 and the second to be 3.

In the specific example of a fragment of permutation problem (lines 6–8), two parameters are declared: \textit{length} defines the size of the permutation, \textit{avoidances} is a set of sequences that must be avoided.

The decision variable perm represents a permutation, which is a sequence of integers from 1 to length. The constraint (line 8) ensures that the sequence is injective, meaning each element is unique.

\subsubsection{Constraint Types}

There are several types of constraints that can be applied, depending on the structure of the problem:
\begin{description}
    \item[Iterating over elements (Line 10):] The \textit{forAll} keyword applies a constraint across all elements of the sequence. For example, this syntax might ensure that certain conditions hold for every element between 1 and length.
    \item[Matrix constraints (Line 11):] In more complex problems, one can quantify over a matrix domain. This is useful for considering all permutations of a particular length, for example.
    \item[Conditional constraints (Line 12):] Conditional constraints apply to a subset of the domain, based on certain conditions. For example, one can restrict constraints to cases where a specific condition is true.
    \item[Pattern avoidance (Line 13):] One can iterate over a set of patterns (in this case, avoidances) and enforce constraints that prevent the sequence from containing any of these patterns.
    \item[Fixed values (Line 14):] Lastly, constraints can assign fixed values to certain positions in the sequence. In this example, the first two positions of perm are assigned values of 2 and 3, respectively. This operation corresponds to permutation application when the sequence is viewed as a permutation.
\end{description}
In all of these examples the \textit{forAll} quantifier can be replaced with \textit{exists} as well, to achieve existential quantification as opposed to universal.

In \Cref{lst:exclav} we give a complete example constraint model, where the parameters values are defined within the model itself for simplicity. This problem is that of finding all permutations of length 4 that classically contain the pattern permutation $21$. We will explain the model in more detail in \Cref{sec:lib}.

\begin{lstlisting}[ language=essence,caption=Essence code which represents classic containment,label=lst:exclav]
letting length be 4
letting classic_containment be {sequence(2,1)}

find perm : sequence (size length, injective) of int(1..length)

such that
    forAll pattern in classic_containment .
         (exists ix : matrix indexed by [int(1..|pattern|)] of int(1..length) .
            (forAll i,j : int(1..|pattern|) . i < j -> ix[i] < ix[j]) /\
            (forAll n1, n2 : int(1..|pattern|) , n1 < n2 .
                pattern(n1) < pattern(n2) <-> perm(ix[n1]) < perm(ix[n2])))
\end{lstlisting}

This model (\Cref{lst:exclav}) can be solved to obtain either a single solution, a specified number of solutions, or all possible solutions. In \Cref{lst:output} we present the result when asking for 10 solutions.

\begin{lstlisting}[ language=essence,caption={Output from running a constraint solver over the 21 classical containment model.},label=lst:output]
# A single, random permutation of length 4, containing 21
{"perm": [1, 2, 4, 3]}


# 10 random permutations of length 4, containing 21
[{"perm": [1, 2, 4, 3]}, {"perm": [1, 3, 2, 4]}, {"perm": [1, 3, 4, 2]}, {"perm": [1, 4, 2, 3]}, 
 {"perm": [1, 4, 3, 2]}, {"perm": [2, 1, 3, 4]}, {"perm": [2, 1, 4, 3]}, {"perm": [2, 3, 1, 4]}, 
 {"perm": [2, 3, 4, 1]}, {"perm": [2, 4, 1, 3]}]

\end{lstlisting}

The strength of constraint modelling lies in its approach to solving problems with multiple constraints. Unlike traditional methods, which typically generate a set of permutations and then filter them based on additional constraints, constraint solvers work by applying all constraints at once during the search process. This direct approach ensures efficiency and avoids unnecessary intermediate steps, making the search for valid solutions more streamlined. With some care in how parameters and variables are named, different constraints can be combined defined in a modular fashion and composed seamlessly, allowing the solver to find solutions that satisfy all conditions simultaneously.

The problem specification language Essence is particularly well-suited for modelling permutation pattern problems due to its expressive power and flexibility in defining complex combinatorial problems declaratively. It allows for a natural representation of sequences, set operations, and matrix indexing, which are fundamental in defining and manipulating permutations. The ability to declare injective sequences and easily handle universal or existential quantification over sets of constraints makes Essence an ideal choice for modelling pattern containment and avoidance problems. Furthermore, Essence's high-level abstraction allows users to focus on formulating the problem without needing to worry about implementation details, which are handled during the translation process into solver-compatible formats. This makes it an efficient tool for exploring and solving permutation pattern problems.

\subsection{Constraint Translation}

After modelling a problem in Essence, the next step is translating the high-level specification into a lower-level representation that can be handled by a solver. This translation process is automatically performed by Conjure and Savile Row~\cite{nightingale2017automatically}, which collectively convert Essence models into a form suitable for constraint solvers. The translation bridges the abstract, declarative model with the computational techniques necessary for solving it.

Typically, the solver chosen by default (e.g., Minion~\cite{minion} for constraint problems) is sufficient for most tasks, but users can opt for other solvers based on the problem's nature. For example, Chuffed~\cite{chu2018chuffed} is another constraint solver, while MiniSAT~\cite{een2003extensible} is used for Boolean satisfiability (SAT) problems. Each solver offers different strategies and algorithms for navigating the solution space, and their suitability depends on the problem at hand.

What is key to note is that the user does not need to concern themselves with this translation process. They only need to focus on modelling the problem in a high-level language like Essence. The translation step transforms the model into formats that can be executed by various solvers, making it efficient and easy to experiment with different solvers without needing to rewrite the core model.

By automating this process, Essence and Conjure significantly reduce errors that may occur during manual translation, while also enabling the flexibility to leverage a range of solvers for a single problem. This ensures that the model remains solver-agnostic, allowing the user to concentrate on the conceptual aspects of the problem rather than the technicalities of the solving process.

\subsection{Constraint Solving}

Once the model has been translated into a solver-compatible format, the solver takes over to search for solutions using a combinatorial search algorithm, most commonly a form of backtrack search. This process systematically assigns values to variables while checking that all constraints are satisfied.

A solver processes the entire model simultaneously, as seen in \Cref{lst:exclav}. Unlike traditional algorithms in permutation patterns, which may generate all permutations or subsets first and then apply filters based on constraints, constraint solvers apply the constraints immediately during the search. This approach avoids unnecessary computation and ensures that only valid solutions are considered.

The search procedure typically works as follows:

\begin{description}
    \item[Variable-Value Assignment:] The solver selects a variable and assigns a value from its domain (the range of possible values). The choice of which variable to assign next can be random, based on a heuristic, or follow a predefined ordering. This is only done after constraint propagation, to avoid evaluating values that are shown to be impossible.
    \item[Constraint Checking and Propagation:] At the beginning of the search and after every time a value is assigned to a variable, the solver checks whether the assignment violates any constraints. If not, the solver propagates this assignment, which involves updating the domains of other variables to reflect the consequences of the assignment. For example, assigning a value to one variable might reduce the possible values for others.
    \item[Backtracking:] If a constraint is violated at any point during the search, the solver backtracks. It undoes the most recent variable assignment and tries a different value, continuing this process until a valid solution is found or all possibilities are exhausted.
    \item[Solution Discovery and Continuation:] When all variables have valid assignments that satisfy the constraints, the solver has found a solution. If multiple solutions are needed, the solver backtracks to explore different combinations of variable assignments, continuing the search to find additional solutions.
\end{description}

The search process is highly efficient because the solver only explores variable assignments that are consistent with the constraints, avoiding unnecessary work. Propagation, in particular, helps to narrow down the search space by eliminating infeasible options early.

In summary, constraint solvers combine trial and error, propagation, and backtracking to efficiently navigate the search space. While this paper focuses on providing models in the Essence language, these models can be solved by a variety of solvers using similar search strategies to find valid solutions.

Having established the fundamental aspects of constraint programming, from modelling to solving, we now turn our attention to a more specialised domain: permutation pattern problems. Permutation patterns are a rich source of combinatorial problems, where one seeks to identify or avoid specific patterns within permutations. The flexibility and power of the Essence language, combined with constraint programming, allow for an elegant and expressive representation of these problems. In the following section, we present a comprehensive library of models specifically designed for permutation pattern problems, showcasing how various types of pattern containment, avoidance, and permutation properties can be encoded within the CP framework. These models form a foundation for exploring and solving a wide array of permutation-related challenges.

\section{Library of Models}
\label{sec:lib}

We will consider a \emph{permutation} $\sigma$ to be an arrangement of the set $\{1, 2, \ldots , n\}$ for some $n\in\mathbb{N}$.
The size of the set a permutation $\sigma$ is defined on is called the length of the permutation and is denoted $|\sigma|$.
$S_n$ is used to denote the set of all permutations of length $n$.
$\sigma$ can be viewed as a bijective function, where $\sigma(i)$ denotes the $i$th member of the permutation and $\sigma^{-1}(j)$ gives the index where $j$ occurs in the permutation.

Two permutations $\pi = \pi(1),\ldots,\pi(n)$ and $\sigma = \sigma(1),\ldots,\sigma(n)$ of the same length are said to be \emph{order isomorphic} \cite{DBLP:journals/dm/Atkinson99} if $\forall i, j, \pi(i) \leq \pi (j)$ if and only if $\sigma(i) \leq \sigma(j)$. The same property applies for sequences as well.

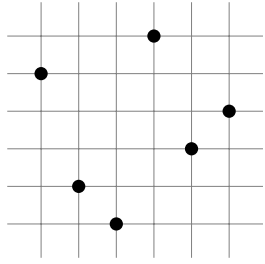
\begin{figure}
\begin{center}
\begin{tikzpicture}[scale=0.5]
\draw [help lines] (0.1,0.1) grid (6.9,6.9);
\fill (1,5) circle (5pt);
\fill (2,2) circle (5pt);
\fill (3,1) circle (5pt);
\fill (4,6) circle (5pt);
\fill (5,3) circle (5pt);
\fill (6,4) circle (5pt);
\end{tikzpicture}
\caption{Plot representation of the permutation $521634$. The x-axis represents the indexes of the permutation, and the y-axis represents the values. Reading left to right, the location of the dot in the $i$th column represents the $i$th value in the permutation.}
\label{fig:permex}
\end{center}
\end{figure}

We will represent permutations by giving the arrangement of the set (often called sequence notation) and as a permutation plot.
\Cref{fig:permex} gives an example of a permutation plot.

The rest of this Section defines the patterns, properties and statistics which we have currently implemented.
These terms are summarised in \Cref{tab:contri}.

All CP models from the upcoming sections with definitions can be found in a supplementary repository~\cite{ozgur_akgun_2023_10215929}. 
This repository contains executable CP models, sample parameter files, raw data for our computational experiments and scripts that can be used to fully rerun the experiments. In addition, the repository contains a Jupyter notebook that can be used to interactively run model fragments through Conjure. The notebook also contains the example scenarios from \Cref{sec:composability}.
The aim of the models is to be as close to the mathematical definitions as possible.

\begin{table}
\begin{center}
\begin{tabular}{c|c|c}
\multicolumn{2}{c|}{Conditions} & Statistics\\ \hline
Pattern Avoidance/Containment & Properties & Number of\\ \hline
Classic &  Simple & Inversions\\
Vincular & Plus-Decomposable & Ascents \\
Bivincular & Minus-Decomposable & Descents \\
Mesh & Blockwise Simple &  Excedances \\
Boxed Mesh & Derangement & Major Index\\
Consecutive & Non-Derangement & \\
 & Involution & \\
 & Parity & \\
\end{tabular}
\caption{Table of all permutation patterns, properties and statistics modelled\label{tab:contri}}
\end{center}
\end{table}

\subsection{Pattern types\label{sec:pattypes}}
A permutation pattern identifies a subsequence of a permutation which satisfies a list of constraints.
If a subsequence of a permutation satisfies the requirements of a given permutation pattern, it is said to \emph{involve} or \emph{contain} the pattern.
If no part of a permutation matches a permutation pattern then it \emph{avoids} the pattern.
There are many types of permutation patterns, in this section we introduce the ones considered in this paper.
We will denote the set of permutations that contains a set of patterns $P$ as Co$(P)$ and the set of permutations which avoid a set of patterns $P$ as Av($P$).

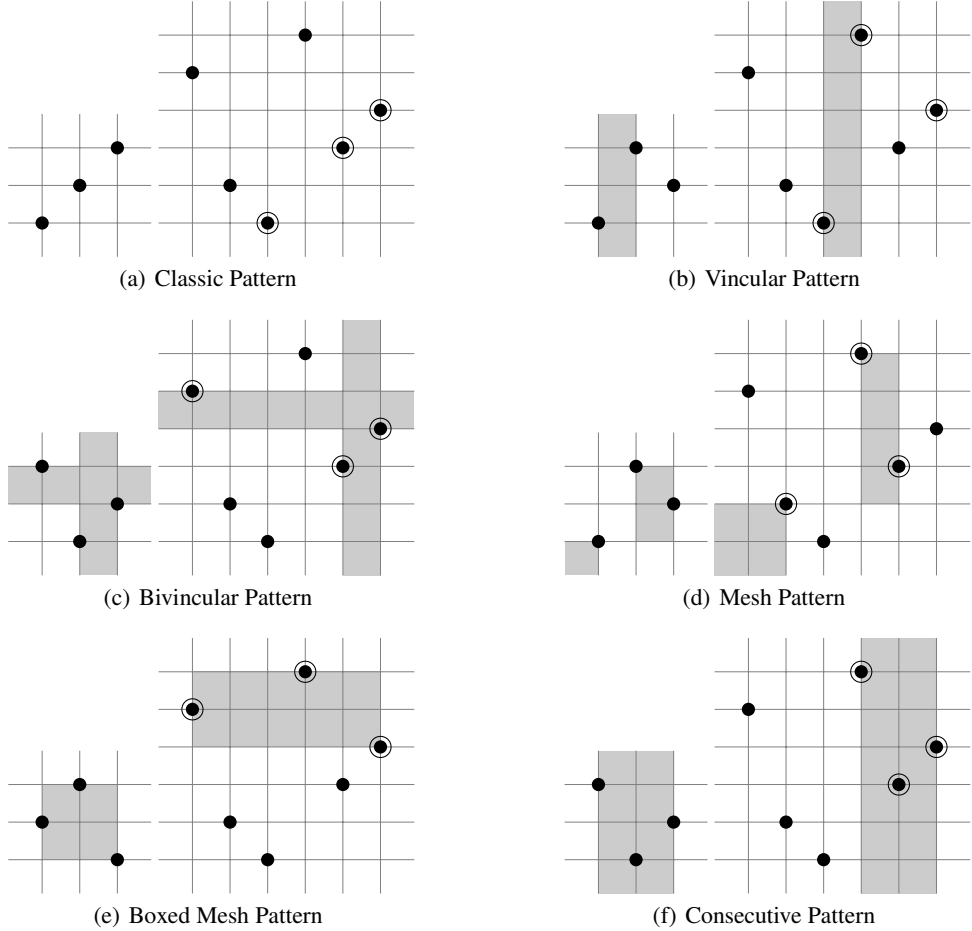
\begin{figure}
\centering
\subfigure[Classic Pattern\label{fig:classcont}]{
\begin{tikzpicture}[scale=0.5]
  \draw [help lines] (0.1,0.1) grid (3.9,3.9);
  \fill (1,1) circle (5pt);
  \fill (2,2) circle (5pt);
  \fill (3,3) circle (5pt);
\end{tikzpicture}
\begin{tikzpicture}[scale=0.5]
  \draw [help lines] (0.1,0.1) grid (6.9,6.9);
  \fill (1,5) circle (5pt);
  \fill (2,2) circle (5pt);
  \fill (3,1) circle (5pt);
  \fill (4,6) circle (5pt);
  \fill (5,3) circle (5pt);
  \fill (6,4) circle (5pt);
\draw (3,1) circle (8pt);
\draw (5,3) circle (8pt);
\draw (6,4) circle (8pt);
\end{tikzpicture}}
\hfil
\subfigure[Vincular Pattern\label{fig:vinccont}]{
\begin{tikzpicture}[scale=0.5]
  \fill[gray!40] (1,0.1) rectangle (2,3.9);
  \draw [help lines] (0.1,0.1) grid (3.9,3.9);
  \fill (1,1) circle (5pt);
  \fill (2,3) circle (5pt);
  \fill (3,2) circle (5pt);
\end{tikzpicture}
\begin{tikzpicture}[scale=0.5]
  \fill[gray!40] (3,0.1) rectangle (4,6.9);
  \draw [help lines] (0.1,0.1) grid (6.9,6.9);
  \fill (1,5) circle (5pt);
  \fill (2,2) circle (5pt);
  \fill (3,1) circle (5pt);
  \fill (4,6) circle (5pt);
  \fill (5,3) circle (5pt);
  \fill (6,4) circle (5pt);
\draw (3,1) circle (8pt);
\draw (4,6) circle (8pt);
\draw (6,4) circle (8pt);
\end{tikzpicture}}

\centering
\subfigure[Bivincular Pattern\label{fig:bivinccont}]{
\begin{tikzpicture}[scale=0.5]
  \fill[gray!40] (2,0.1) rectangle (3,3.9);
  \fill[gray!40] (0.1,2) rectangle (3.9,3);
  \draw [help lines] (0.1,0.1) grid (3.9,3.9);
  \fill (1,3) circle (5pt);
  \fill (2,1) circle (5pt);
  \fill (3,2) circle (5pt);
\end{tikzpicture}
\begin{tikzpicture}[scale=0.5]
  \fill[gray!40] (5,0.1) rectangle (6,6.9);
  \fill[gray!40] (0.1,4) rectangle (6.9,5);
  \draw [help lines] (0.1,0.1) grid (6.9,6.9);
  \fill (1,5) circle (5pt);
  \fill (2,2) circle (5pt);
  \fill (3,1) circle (5pt);
  \fill (4,6) circle (5pt);
  \fill (5,3) circle (5pt);
  \fill (6,4) circle (5pt);
\draw (1,5) circle (8pt);
\draw (5,3) circle (8pt);
\draw (6,4) circle (8pt);
\end{tikzpicture}}
\hfil
\subfigure[Mesh Pattern\label{fig:meshcont}]{
\begin{tikzpicture}[scale=0.5]
  \fill[gray!40] (0.1,0.1) rectangle (1,1);
  \fill[gray!40] (2,2) rectangle (3,3);
  \fill[gray!40] (2,1) rectangle (3,2);
  \draw [help lines] (0.1,0.1) grid (3.9,3.9);
  \fill (1,1) circle (5pt);
  \fill (2,3) circle (5pt);
  \fill (3,2) circle (5pt);
\end{tikzpicture}
\begin{tikzpicture}[scale=0.5]
  \fill[gray!40] (0.1,0.1) rectangle (2,2);
  \fill[gray!40] (4,2) rectangle (5,6);
  \draw [help lines] (0.1,0.1) grid (6.9,6.9);
  \fill (1,5) circle (5pt);
  \fill (2,2) circle (5pt);
  \fill (3,1) circle (5pt);
  \fill (4,6) circle (5pt);
  \fill (5,3) circle (5pt);
  \fill (6,4) circle (5pt);
\draw (2,2) circle (8pt);
\draw (4,6) circle (8pt);
\draw (5,3) circle (8pt);
\end{tikzpicture}}

\centering
\subfigure[Boxed Mesh Pattern\label{fig:bmeshcont}]{
\begin{tikzpicture}[scale=0.5]
  \fill[gray!40] (1,1) rectangle (3,3);
  \draw [help lines] (0.1,0.1) grid (3.9,3.9);
  \fill (1,2) circle (5pt);
  \fill (2,3) circle (5pt);
  \fill (3,1) circle (5pt);
\end{tikzpicture}
\begin{tikzpicture}[scale=0.5]
  \fill[gray!40] (1,4) rectangle (6,6);
  \draw [help lines] (0.1,0.1) grid (6.9,6.9);
  \fill (1,5) circle (5pt);
  \fill (2,2) circle (5pt);
  \fill (3,1) circle (5pt);
  \fill (4,6) circle (5pt);
  \fill (5,3) circle (5pt);
  \fill (6,4) circle (5pt);
\draw (1,5) circle (8pt);
\draw (4,6) circle (8pt);
\draw (6,4) circle (8pt);
\end{tikzpicture}}
\hfil
\subfigure[Consecutive Pattern\label{fig:conseccont}]{
\begin{tikzpicture}[scale=0.5]
  \fill[gray!40] (1,0.1) rectangle (3,3.9);
  \draw [help lines] (0.1,0.1) grid (3.9,3.9);
  \fill (1,3) circle (5pt);
  \fill (2,1) circle (5pt);
  \fill (3,2) circle (5pt);
\end{tikzpicture}
\begin{tikzpicture}[scale=0.5]
  \fill[gray!40] (4,0.1) rectangle (6,6.9);
  \draw [help lines] (0.1,0.1) grid (6.9,6.9);
  \fill (1,5) circle (5pt);
  \fill (2,2) circle (5pt);
  \fill (3,1) circle (5pt);
  \fill (4,6) circle (5pt);
  \fill (5,3) circle (5pt);
  \fill (6,4) circle (5pt);
\draw (4,6) circle (8pt);
\draw (5,3) circle (8pt);
\draw (6,4) circle (8pt);
\end{tikzpicture}}

\caption{Examples of all patterns and for each an example containment in a larger permutation.
In each subfigure the pattern is on the left and an occurrence of the pattern is indicated in the right permutation.
We highlight a classic occurrence (the elements that are order isomorphic to the pattern permutation) with circles around the nodes.
All additional constraints (in terms of adjacency in index or values, or the forbidden cells/regions) are shaded.
In the target permutation the shaded regions scale with respect to the elements which are representative of the classic pattern.}
\label{fig:pattex}
\end{figure}

We say that a permutation $\pi = \pi(1)\ldots\pi(k)$ is \emph{classically contained} in a permutation $\sigma = \sigma(1)\ldots\sigma(m)$, where $k \leq m$, if there is a subsequence $\sigma(i_1)\ldots \sigma(i_k)$ in $\sigma$ that is order isomorphic to $\pi$. 
For example, the permutation $\pi=123$ can be found in $\sigma=521634$ as the order isomorphic subsequence $134=\sigma(3)\sigma(5)\sigma(6)$.
Being \emph{classically contained} is a partial order on the set of permutations \cite{brignall2010survey}.
In \Cref{fig:classcont} on the left is the pattern permutation, on the right we have highlighted an occurrence of the classical pattern by circling the elements.

\Cref{lst:classiccont} represents the model for classic containment, where in Lines 1 and 2 we limit the model to finding permutations of length 4 and limit the set of permutation patterns that we look for to the just the permutation $21$.
\begin{lstlisting}[ language=essence,caption=Essence code which represents classic containment,label=lst:classiccont]
letting length be 4
letting classic_containment be {sequence(2,1)}

find perm : sequence (size length, injective) of int(1..length)

such that
$ For each pattern in the classic_containment set we will look for at least one occurence of it
    forAll pattern in classic_containment .
$ We use a matrix to represent the permutation in two dimensions, this is not really needed for classic patterns, 
$ but to allow for the code of the different patterns to be composable we have added this here as well. 
         (exists ix : matrix indexed by [int(1..|pattern|)] of int(1..length) .
$ We now look constraint perm to contain an order isomorphic copy of pattern. 
            (forAll i,j : int(1..|pattern|) . i < j -> ix[i] < ix[j]) /\
            (forAll n1, n2 : int(1..|pattern|) , n1 < n2 .
                pattern(n1) < pattern(n2) <-> perm(ix[n1]) < perm(ix[n2])))
\end{lstlisting}

\emph{Vincular patterns} (introduced by  \cite{babson2000generalized}) specify adjacency conditions.
Let $\pi=\pi(1) \ldots \pi(k)$, $\sigma=\sigma(1)\ldots \sigma(m)$ and let $A \subseteq \{0,\ldots,k\}$. To simplify the pattern definition, we define $\pi(0)=0$ and $\pi(k+1)=k+1$.
An occurrence of the vincular pattern $(\pi, A)$ in $\sigma$ is a subsequence $\sigma(i_1)\ldots \sigma(i_k)$ of $\sigma$ such that
$\sigma(i_1)\ldots\sigma(i_k)$ is an occurrence of $\pi$ in the classical sense, and $\forall a \in A.\ i_{a+1}=i_{a} +1$.
We call $A$ the set of \emph{adjacencies}. The code in \Cref{lst:vincularcont} shows the code for any given length and looking for the vincular pattern $(132, \{1\})$.

For example, the vincular pattern $(132,\{1\})$ can be found in $\sigma=521634$ as the subsequence $164=\sigma(3)\sigma(4)\sigma(6)$.
\Cref{fig:vinccont} shows an example of a vincular pattern by showing the pattern with the order isomorphic subsequence and highlights the adjacency requirement by shading the column between the indices both in the pattern and the permutation which contains the pattern.

\begin{lstlisting}[ language=essence,caption=Essence code which represents vincular containment,label=lst:vincularcont]
given length : int
letting vincular_containment be  { [[1,3,2], [1]] } 

find perm : sequence (size length, injective) of int(1..length)

such that
$ We check each vincular pair which consists of the permutation (pattern) and adjacency set (bar).
    forAll (pattern, bars) in vincular_containment .
        exists ix : matrix indexed by [int(1..|pattern|)] of int(1..length) .
$ First we look for a classic pattern occurence
            (forAll i,j : int(1..|pattern|) . i < j -> ix[i] < ix[j]) /\
            (forAll n1, n2 : int(1..|pattern|) , n1 < n2 .
                (pattern(n1) < pattern(n2) <-> perm(ix[n1]) < perm(ix[n2]))) .
$ Then we look in the matrix representation if the adjacency for this pattern is not violated.      
            (forAll bar in bars . ix[bar] + 1 = ix[bar+1])
\end{lstlisting}

\emph{Bivincular patterns} (as introduced in \cite{bousquet20102}) are vincular patterns which have an additional set defining which values have to be adjacent.

More formally, an occurrence of the bivincular pattern $(\pi, A, B)$ in $\sigma$, with $A,B \subseteq \{0,\ldots,k\}$ and $|\pi|=k$, is a subsequence $\sigma(i_1)\ldots \sigma(i_k)$ of $\sigma$ such that the following all hold:
\begin{itemize}
    \item $\sigma(i_1)\ldots\sigma(i_k)$ is an occurrence of $\pi$ in the classical sense,
    \item $\forall a\in A.\ i_{a+1}=i_{a}+1$
    \item $\forall b\in B.\ j_{b+1} = j_{b}+1$
\end{itemize}
where $\{\sigma(i_1),\ldots,\sigma(i_k)\} = \{j_1,\ldots,j_k\}$ and $j_1 < j_2 < \cdots < j_k$. By convention $i_0=j_0=0$ and $i_{k+1}=j_{k+1}=n+1$.
As with vincular patterns, $A$ is the set of adjacencies of indexes, but now the set $B$ defines the adjacencies of values. 
In \Cref{lst:bivincont} we show the model looking for permutations of a given length containing the set of bivincular patterns.

For example, the bivincular pattern $(312,\{2\},\{2\})$ can be found in $\sigma=521634$ as the subsequence $534=\sigma(1)\sigma(5)\sigma(6)$.
\Cref{fig:bivinccont} illustrates this bivincular containment with the pattern permutation and the shading of the adjacency in values (rows) and indices (columns) highlighted.

\newpage
\newpage
\newpage

\begin{lstlisting}[ language=essence,caption=Essence code which represents bivincular containment,label=lst:bivincont,float, floatplacement=T]
given length : int
given bivincular_containment : set of (sequence (injective) of int, set of int, set of int)

find perm : sequence (size length, injective) of int(1..length)

$ Creating a padded version of perm, where position 0 contains the value 0 and position length+1 contains the value length+1, 
$ to be able to follow the convention.
find permPadded : matrix indexed by [int(0..length+1)] of int(0..length+1)
such that permPadded[0] = 0, permPadded[length+1] = length+1
such that forAll i : int(1..length) . permPadded[i] = perm(i)

such that
$ We look for each bivincular pattern consisting of the permutation, the index adjacencies and the value adjacencies.
    forAll (pattern, ind_bars, val_bars) in bivincular_containment .
$ Build the inverse of 'pattern'. This is completely evaluated before solving.
        exists patterninv: matrix indexed by [int(0..|pattern|+1)] of int(0..|pattern|+1),
            patterninv[0] = 0 /\ patterninv[|pattern|+1] = |pattern|+1 /\
            (forAll i: int(1..|pattern|) . patterninv[pattern(i)] = i).

$ Look for all places where the pattern can occur classically
        exists ix : matrix indexed by [int(0..|pattern|+1)] of int(0..length+1),
            and([ ix[0]=0, ix[|pattern|+1]=length+1 ,
                forAll i : int(0..|pattern|) . ix[i] < ix[i+1] ,
                forAll n1, n2 : int(1..|pattern|) , n1 < n2 .
                    pattern(n1) < pattern(n2) <-> permPadded[ix[n1]] < permPadded[ix[n2]]
                ]) .
$ Check if the index adjacency is not violated
            ((forAll bar in ind_bars . ix[bar] + 1 = ix[bar+1])
            /\
$ And check if the value adjacency is not violated
            (forAll bar in val_bars . permPadded[ix[patterninv[bar]]]+1 = 
                    permPadded[ix[patterninv[bar+1]]]))
\end{lstlisting}

\begin{lstlisting}[ language=essence,caption=Essence code which represents mesh containment,label=lst:meshcont,float, floatplacement=T]
given length : int
given mesh_containment : set of (sequence(injective) of int, relation of (int * int))

find perm : sequence (size length, injective) of int(1..length)

$ Creating a padded version of perm, where position 0 contains the value 0 and position length+1 contains the value length+1, 
$ to be able to follow the convention.
find permPadded : matrix indexed by [int(0..length+1)] of int(0..length+1)
such that permPadded[0] = 0, permPadded[length+1] = length+1
such that forAll i : int(1..length) . perm(i) = permPadded[i]

such that
$ pattern is the pattern, mesh is the mesh as a relation
    forAll (pattern, mesh) in mesh_containment .
$ Build the inverse of 'pattern'. This is completely evaluated before solving.
    exists patterninv: matrix indexed by [int(0..|pattern|+1)] of int(0..|pattern|+1),
                patterninv[0] = 0 /\ patterninv[|pattern|+1] = |pattern|+1 /\
                (forAll i: int(1..|pattern|) . patterninv[pattern(i)] = i).
$ Look for all places where the pattern can classically occur
        exists ix : matrix indexed by [int(0..|pattern|+1)] of int(0..length+1),
            and([ ix[0]=0, ix[|pattern|+1]=length+1 , 
                forAll i : int(0..|pattern|) . ix[i] < ix[i+1] , 
                forAll n1, n2 : int(1..|pattern|) , n1 < n2 .
                    pattern(n1) < pattern(n2) <-> permPadded[ix[n1]] < permPadded[ix[n2]]
                ]) .
$ If we find the pattern, make sure there is NOT at least one value in some cell of the mesh
            !( exists (i,j) in mesh.
               exists z: int(ix[i]+1..ix[i+1]-1). (permPadded[ix[patterninv[j]]] <= permPadded[z] /\ 
                        permPadded[z] <= permPadded[ix[patterninv[j+1]]]) )
\end{lstlisting}

All of the above patterns can be generalised as \emph{mesh patterns}, which were introduced in \cite{branden2011mesh}.
A mesh pattern of length $k$ is a pair $(\pi, R)$ with $\pi \in S_k$ and $R \subseteq [0, k] \times [0, k]$, a set of pairs of integers.
The elements of $R$ identify the lower left corners of unit squares in the plot of $\pi$ and specify forbidden regions.
An occurrence of a mesh pattern $(\pi, R)$ in a permutation $\sigma$ is a subsequence $\sigma(i_1)\ldots \sigma(i_k)$ such that the following holds
\begin{itemize}
    \item $\sigma(i_1)\ldots \sigma(i_k)$ is order isomorphic to $\pi$
    \item $(x,y)\in R \Rightarrow$ there does not exist $i\in \{1,\ldots,n\} : i_{x} < i < i_{x+1} \land \sigma(i_{\pi^{-1}(y)}) < \sigma(i) < \sigma(i_{\pi^{-1}(y+1)})$.
\end{itemize}
For this definition we extend the subsequence $i_j$, $\sigma$ and $\pi$ with $i_0=0$, $i_{k+1} = n+1$, $\pi(0)=0$, $\pi(k+1)=k+1$, $\sigma(0)=0$ and $\sigma(n+1)=n+1$. 
The extra terms for $i_j$, $\sigma$ and $\pi$ in the second part of the definition above allow mesh patterns to constrain a permutation outside of the pattern.
The model of mesh containment (together with the extended mesh) is in \Cref{lst:meshcont}.
An example of a mesh pattern in $\sigma=521634$ is $(132,\{(0,0),(2,1),(2,2)\})$, which can be found as the subsequence $263=\sigma(2)\sigma(4)\sigma(5)$.
This example is illustrated in \Cref{fig:meshcont}, where each of the forbidden regions/squares is shaded.

A \emph{boxed mesh pattern} (or \emph{boxed pattern}, as introduced in \cite{boxedmesh}), is a special case of a mesh pattern $P=(\pi, R)$ where $\pi$ is a permutation of length $k$ and $R=[1,k-1]\times[1,k-1]$.
$P$ is then denoted by \framebox{$\pi$}.
For example the boxed pattern \framebox{$231$} is contained in the permutation $236514$ as the subsequence $351=\sigma(2)\sigma(4)\sigma(5)$.
\Cref{fig:bmeshcont} shows this example, and illustrates that for a boxed mesh pattern the box inside the permutation is the forbidden region.

\emph{Consecutive patterns} are a special case of vincular patterns, where it is necessary that \emph{all} entries are adjacent.
For example, the consecutive pattern $(312,\{1,2\})$ can be found inside $152463$ as the subsequence $524=\sigma(2)\sigma(3)\sigma(4)$.
This example is shown in \Cref{fig:conseccont}.

We say a permutation $\sigma$ \emph{avoids} any of the above pattern types, if the permutation $\pi$ of the pattern is classically avoided in $\sigma$, or it is classically contained in $\sigma$ but for every occurrence of the classical pattern the additional constraints on indices or values are not upheld.

\begin{lstlisting}[ language=essence,caption=Essence code which represents classic avoidance,label=lst:classicav,float, floatplacement=T]
given length : int 
given classic_avoidance : set of sequence of int

find perm : sequence (size length, injective) of int(1..length)

such that
$ For each pattern in the classic_avoidance set we will look for at least one occurence of it
    forAll pattern in classic_avoidance .
$ We use a matrix to represent the permutation in two dimensions, this is not really needed for classic patterns, 
$ but to allow for the code of the different patterns to be composable we have added this here as well. 
$ While the classic pattern constraint stays the same we are now negating the whole thing (see the ! before the exists statement)
$ So we are looking for a permutation in which it is not possible to find the pattern in.
         !(exists ix : matrix indexed by [int(1..|pattern|)] of int(1..length) .
$ We now look constraint perm to contain an order isomorphic copy of pattern. 
            (forAll i,j : int(1..|pattern|) . i < j -> ix[i] < ix[j]) /\
            (forAll n1, n2 : int(1..|pattern|) , n1 < n2 .
                pattern(n1) < pattern(n2) <-> perm(ix[n1]) < perm(ix[n2])))
\end{lstlisting}

\Cref{lst:classicav} shows the model for classic pattern avoidance. We can see that in comparison to classic containment (\Cref{lst:classiccont}) there is a negation sign (denoted by !) before the matrix which defines the pattern constraint.

Such simple negations apply through all the models, for example \Cref{lst:meshav} where if there is an occurrence of the pattern there are points that violate the mesh regions.

\begin{lstlisting}[ language=essence,caption=Essence code which represents mesh avoidance,label=lst:meshav,float, floatplacement=T]
given length : int
given mesh_avoidance : set of (sequence(injective) of int, relation of (int * int))

find perm : sequence (size length, injective) of int(1..length)

$ Creating a padded version of perm, where position 0 contains the value 0 and position length+1 contains the value length+1, 
$ to be able to follow the convention.
find permPadded : matrix indexed by [int(0..length+1)] of int(0..length+1)
such that permPadded[0] = 0, permPadded[length+1] = length+1
such that forAll i : int(1..length) . perm(i) = permPadded[i]

such that
$ pattern is the pattern, mesh is the mesh as a relation
    forAll (pattern, mesh) in mesh_avoidance .
$ Build the inverse of 'pattern'. This is completely evaluated before solving.
    exists patterninv: matrix indexed by [int(0..|pattern|+1)] of int(0..|pattern|+1),
                patterninv[0] = 0 /\ patterninv[|pattern|+1] = |pattern|+1 /\
                (forAll i: int(1..|pattern|) . patterninv[pattern(i)] = i).
$ Look for all places where the pattern can occur, in each we need to violate the mesh.
$ This changes from an 'exists' to a 'for all'
        forAll ix : matrix indexed by [int(0..|pattern|+1)] of int(0..length+1),
            and([ ix[0]=0, ix[|pattern|+1]=length+1, forAll i : int(0..|pattern|) . ix[i] < ix[i+1]
                , forAll n1, n2 : int(1..|pattern|) , n1 < n2 .
                    pattern(n1) < pattern(n2) <-> permPadded[ix[n1]] < permPadded[ix[n2]]
                ]) .
$ If we find the pattern, make sure there is at least one value in some cell of the mesh
            ( exists (i,j) in mesh.
              exists z: int(ix[i]+1..ix[i+1]-1). (permPadded[ix[patterninv[j]]] <= permPadded[z] /\ 
                        permPadded[z] <= permPadded[ix[patterninv[j+1]]]) )

\end{lstlisting}

\begin{figure}
\centering
\subfigure[$12345$ avoids the classic pattern $21$.\label{fig:classav}]{
\begin{tikzpicture}[scale=0.5]
  \draw [help lines] (0.1,0.1) grid (2.9,2.9);
  \fill (1,2) circle (5pt);
  \fill (2,1) circle (5pt);
\end{tikzpicture}
\begin{tikzpicture}[scale=0.5]
  \draw [help lines] (0.1,0.1) grid (5.9,5.9);
  \fill (1,1) circle (5pt);
  \fill (2,2) circle (5pt);
  \fill (3,3) circle (5pt);
  \fill (4,4) circle (5pt);
  \fill (5,5) circle (5pt);
\end{tikzpicture}}
\hfil
\subfigure[$652413$ avoids the mesh pattern $(132,\{(2,0),(2,1),(2,2)\})$.\label{fig:meshav1}]{
\begin{tikzpicture}[scale=0.5]
    \fill[gray!40] (2,2) rectangle (3,3);
    \fill[gray!40] (2,1) rectangle (3,2);
    \fill[gray!40] (2,0) rectangle (3,1);
    \draw [help lines] (0.1,0.1) grid (3.9,3.9);
    \fill (1,1) circle (5pt);
    \fill (2,3) circle (5pt);
    \fill (3,2) circle (5pt);
\end{tikzpicture}
\begin{tikzpicture}[scale=0.5]
  \draw [help lines] (0.1,0.1) grid (5.9,5.9);
  \fill (1,1) circle (5pt);
  \fill (2,2) circle (5pt);
  \fill (3,3) circle (5pt);
  \fill (4,4) circle (5pt);
  \fill (5,5) circle (5pt);
\end{tikzpicture}}

\centering
\subfigure[$652413$ avoids the mesh pattern $(132,\{(2,0),(2,1),(2,2)\})$.\label{fig:meshav2}]{
\begin{tikzpicture}[scale=0.5]
    \fill[gray!40] (2,2) rectangle (3,3);
    \fill[gray!40] (2,1) rectangle (3,2);
    \fill[gray!40] (2,0) rectangle (3,1);
    \draw [help lines] (0.1,0.1) grid (3.9,3.9);
    \fill (1,1) circle (5pt);
    \fill (2,3) circle (5pt);
    \fill (3,2) circle (5pt);
\end{tikzpicture}
\begin{tikzpicture}[scale=0.5]
    \fill[gray!40] (4,0.1) rectangle (6,4);
    \draw [help lines] (0.1,0.1) grid (6.9,6.9);
    \fill (1,6) circle (5pt);
    \fill (2,5) circle (5pt);
    \fill (3,2) circle (5pt);
    \fill (4,4) circle (5pt);
    \fill (5,1) circle (5pt);
    \fill (6,3) circle (5pt);
    \draw (3,2) circle (8pt);
    \draw (4,4) circle (8pt);
    \draw (6,3) circle (8pt);
\end{tikzpicture}}

\caption{Examples of pattern avoidance.
In subfigures (a) and (b) there is no subsequence in the target permutation which is order isomorphic to the pattern permutation (i.e. there is no occurrence of the classic pattern in the target permutation).
Subfigure (c) avoids the mesh pattern even though the pattern permutation is contained classically, in each occurrence (there is only one) there are (at least one) elements in the target permutation which lie in the shaded/forbidden region. \label{fig:avoid}}
\end{figure}
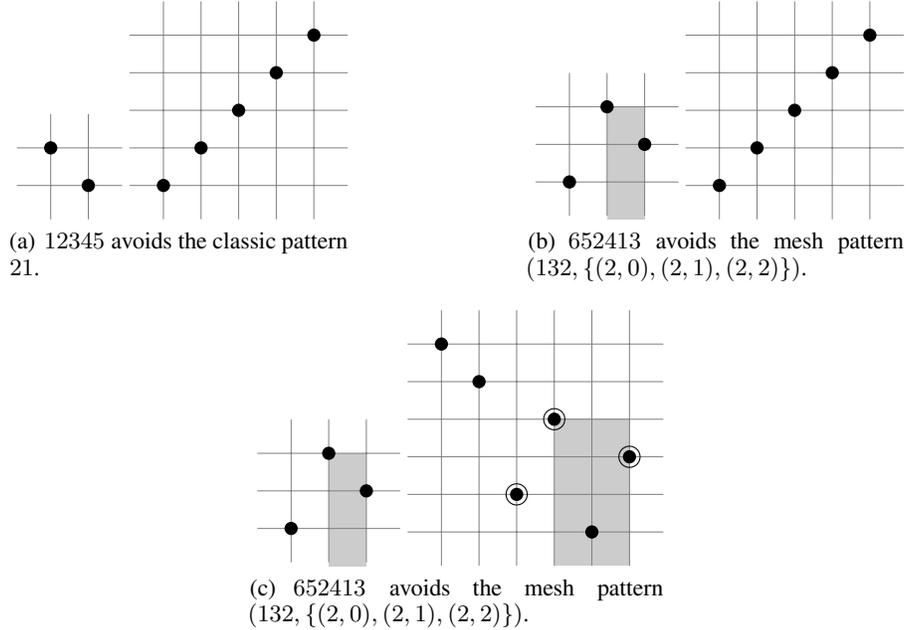

So for example the permutation $\sigma=12345$ avoids the classic pattern $\pi=21$, as there are no occurrences of it.
\Cref{fig:classav} attempts to illustrate this avoidance.
Similarly, the permutation $\sigma=12345$ avoids the mesh pattern $(132, \{(0,0),(2,0),(2,1),(2,2)\})$ as there is no classical occurrence of $132$ in $\sigma$, as can be seen from \Cref{fig:meshav1}.
Finally, $(132, \}(0,0),(2,0),(2,1),(2,2))\}$ is not contained in the permutation $\sigma=652413$ even though there is an occurrence of $132$ in the $\sigma$, namely as $243=\sigma(3)\sigma(4)\sigma(6)$, but there is an element of the permutation that violates the forbidden region.
In \Cref{fig:meshav2} we can see that the element $1=\sigma(5)$ is in the $(2,0)$ shaded/forbidden region.

\subsection{Permutation Properties\label{sec:patprops}}
In addition to containing (or avoiding) patterns in permutations and enumerating these permutations, permutations can have certain structural properties.
We give some of the most common properties here, all of which we will model with CP.
\Cref{tab:contri} contains a list of the properties which we have modelled thus far.
For each property, we can require solutions to either do, or do not, satisfy the property.

\begin{figure}
\centering
\subfigure[Intervals in a permutation\label{fig:inter}]{
\begin{tikzpicture}[scale=0.5]
\draw [help lines] (0.1,0.1) grid (7.9,7.9);
  \fill (1,1) circle (5pt);
  \fill (2,6) circle (5pt);
  \fill (3,3) circle (5pt);
  \fill (4,2) circle (5pt);
  \fill (5,5) circle (5pt);
  \fill (6,4) circle (5pt);
  \fill (7,7) circle (5pt);
\draw[line width=1.5pt] (2.5,1.5) rectangle (4.5,3.5);
\draw[line width=1.5pt] (4.5,3.5) rectangle (6.5,5.5);
\draw[line width=1.5pt] (2.5,1.5) rectangle (6.5,5.5);
\draw[line width=1.5pt] (1.5,1.5) rectangle (6.5,6.5);
\draw[line width=1.5pt] (0.5,0.5) rectangle (6.5,6.5);
\draw[line width=1.5pt] (1.5,1.5) rectangle (7.5,7.5);
\end{tikzpicture}}
\hfil
\subfigure[Plus-decomposable permutation\label{fig:plus}]{
\begin{tikzpicture}[scale=0.5]
\draw [help lines] (0.1,0.1) grid (6.9,6.9);
  \fill (1,2) circle (5pt);
  \fill (2,1) circle (5pt);
  \fill (3,3) circle (5pt);
  \fill (4,6) circle (5pt);
  \fill (5,5) circle (5pt);
  \fill (6,4) circle (5pt);
\draw[line width=1.5pt] (0.5,0.5) rectangle (3.5,3.5);
\draw[line width=1.5pt] (3.5,3.5) rectangle (6.5,6.5);
\end{tikzpicture}}
\hfil
\subfigure[Minus-decomposable permutation\label{fig:minus}]{
\begin{tikzpicture}[scale=0.5]
\draw [help lines] (0.1,0.1) grid (6.9,6.9);
  \fill (1,5) circle (5pt);
  \fill (2,4) circle (5pt);
  \fill (3,6) circle (5pt);
  \fill (4,1) circle (5pt);
  \fill (5,2) circle (5pt);
  \fill (6,3) circle (5pt);
\draw[line width=1.5pt] (0.5,3.5) rectangle (3.5,6.5);
\draw[line width=1.5pt] (3.5,0.5) rectangle (6.5,3.5);
\end{tikzpicture}}
\hfil
\subfigure[Simple permutation\label{fig:simple}]{
\begin{tikzpicture}[scale=0.5]
\draw [help lines] (0.1,0.1) grid (6.9,6.9);
  \fill (1,2) circle (5pt);
  \fill (2,4) circle (5pt);
  \fill (3,6) circle (5pt);
  \fill (4,1) circle (5pt);
  \fill (5,3) circle (5pt);
  \fill (6,5) circle (5pt);
\end{tikzpicture}}

\subfigure[Block-decomposition of a permutation\label{fig:block}]{
\begin{tikzpicture}[scale=0.5]
\draw [help lines] (0.1,0.1) grid (6.9,6.9);
  \fill (1,5) circle (5pt);
  \fill (2,2) circle (5pt);
  \fill (3,1) circle (5pt);
  \fill (4,6) circle (5pt);
  \fill (5,3) circle (5pt);
  \fill (6,4) circle (5pt);
\draw[line width=1.5pt] (0.5,4.5) rectangle (1.5,5.5);
\draw[line width=1.5pt] (1.5,0.5) rectangle (3.5,2.5);
\draw[line width=1.5pt] (3.5,5.5) rectangle (4.5,6.5);
\draw[line width=1.5pt] (4.5,2.5) rectangle (6.5,4.5);
\end{tikzpicture}}
\hfil
\subfigure[Block-wise simple permutation\label{fig:bsimp}]{
\begin{tikzpicture}[scale=0.5]
\draw [help lines] (0.1,0.1) grid (7.9,7.9);
  \fill (1,4) circle (5pt);
  \fill (2,2) circle (5pt);
  \fill (3,5) circle (5pt);
  \fill (4,3) circle (5pt);
  \fill (5,7) circle (5pt);
  \fill (6,1) circle (5pt);
  \fill (7,6) circle (5pt);
\end{tikzpicture}}
\hfil
\subfigure[Not block-wise simple permutation\label{fig:nbsimp}]{
\begin{tikzpicture}[scale=0.5]
\draw [help lines] (0.1,0.1) grid (5.9,5.9);
  \fill (1,2) circle (5pt);
  \fill (2,4) circle (5pt);
  \fill (3,5) circle (5pt);
  \fill (4,1) circle (5pt);
  \fill (5,3) circle (5pt);
\draw[line width=1.5pt] (1.5,3.5) rectangle (3.5,5.5);
\end{tikzpicture}}
\hfil
\subfigure[Involution\label{fig:invo}]{
\begin{tikzpicture}[scale=0.5]
\draw [help lines] (0.1,0.1) grid (4.9,4.9);
  \fill (1,1) circle (5pt);
  \fill (2,2) circle (5pt);
  \fill (3,4) circle (5pt);
  \fill (4,3) circle (5pt);
\end{tikzpicture}}

\centering
\subfigure[Derangement\label{fig:deran}]{
\begin{tikzpicture}[scale=0.5]
\draw [help lines] (0.1,0.1) grid (4.9,4.9);
\draw[red,line width=2pt] (0,0) -- (5,5);
  \fill (1,4) circle (5pt);
  \fill (2,3) circle (5pt);
  \fill (3,1) circle (5pt);
  \fill (4,2) circle (5pt);
\end{tikzpicture}}
\hfil
\subfigure[Non-derangement\label{fig:nond}]{
\begin{tikzpicture}[scale=0.5]
\draw [help lines] (0.1,0.1) grid (4.9,4.9);
\draw[red,line width=2pt] (0,0) -- (5,5);
  \fill (1,2) circle (5pt);
  \fill (2,4) circle (5pt);
  \fill (3,3) circle (5pt);
  \fill (4,1) circle (5pt);
\end{tikzpicture}}
\hfil
\subfigure[Parity\label{fig:parity}]{
\begin{tikzpicture}[scale=0.5]
\draw [help lines] (0.1,0.1) grid (4.9,4.9);
  \fill (1,3) circle (5pt);
  \fill (2,4) circle (5pt);
  \fill (3,1) circle (5pt);
  \fill (4,2) circle (5pt);
\end{tikzpicture}}
\hfil
\subfigure[Non-parity\label{fig:nonp}]{
\begin{tikzpicture}[scale=0.5]
\draw [help lines] (0.1,0.1) grid (4.9,4.9);
  \fill (1,2) circle (5pt);
  \fill (2,4) circle (5pt);
  \fill (3,1) circle (5pt);
  \fill (4,3) circle (5pt);
\end{tikzpicture}}

\caption{Examples permutation properties. Intervals (of length 1 and up to $n$-1, where $n$ is the length of the permutation) and blocks are indicated using squares around the nodes that contain them.
The red diagonal line for the derangement/non-derangement indicates the $\pi(i)=i$ fixed point property. \label{fig:props}}
\end{figure}
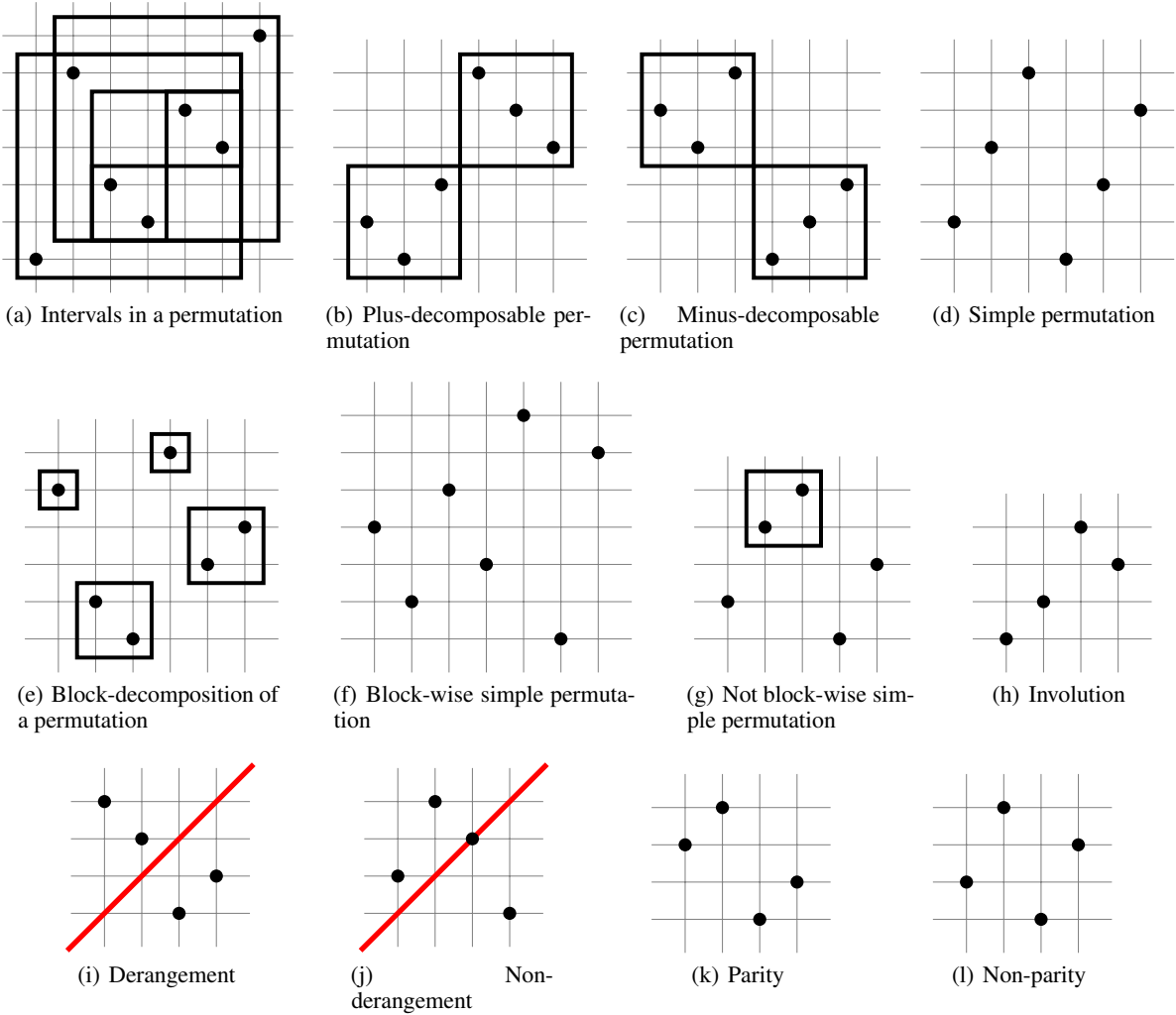

An \emph{interval} of a permutation $\pi$ corresponds to a set of contiguous indices $I = [a, b]$ such that the set of values $\pi(I) = \{\pi(i) : i \in I\}$ is also contiguous.
Every permutation of length $n$ has intervals of lengths 0, 1 and $n$.
If a permutation $\pi$ has no other intervals, then $\pi$ is said to be \emph{simple} \cite{brignall2010survey}.
\Cref{lst:simple} describes the constraints of a simple permutation (in terms of its intervals).
For example, $\pi=1632547$ is not simple as it contains the intervals $\pi(3)\pi(4)$, $\pi(5)\pi(6)$, $\pi(3)\pi(4)\pi(5)\pi(6)$, $\pi(2)\pi(3)\pi(4)\pi(5)\pi(6)$, $\pi(2)\pi(3)\pi(4)\pi(5)\pi(6)\pi(7)$ and $\pi(1)\pi(2)\pi(3)\pi(4)\pi(5)\pi(6)$, as indicated in \Cref{fig:inter} through bold boxes around the intervals.
The permutation $246135$ in \Cref{fig:simple} is simple, as the only intervals it contains are of length 0, 1 or the whole permutation.

\begin{lstlisting}[language=essence,caption=Essence code which represents simple permutations,label=lst:simple]
given length : int
find perm : sequence (size length, injective) of int(1..length)

$ Simple permutations do not contain intervals.
$ An interval is a set of contiguous indices such that the values are also contiguous. 
$ perm is a simple permutation, subperm is a set of contiguous indices.
such that
   and ( [ max(subperm) - min(subperm) + 1 != |subperm| $ the values are not contiguous
    | i : int(1..length)                                $ start index of the sub perm
    , j : int(1..length)                                $ end index of the sub perm
    , i < j                                             $ start comes before end
    , (i,j) != (1,length)                               $ the subpermutation is not the full permutation
    , letting subperm be [perm(k) | k : int(i..j)]      $ give the sub perm a name
    ])
\end{lstlisting}

Given a permutation $\sigma=\sigma(1)\ldots\sigma(m)$ of length $m$ and non-empty permutations $\alpha_{1},\ldots,\alpha_{m}$ the \emph{inflation} of $\sigma$ by $\alpha_{1},\ldots,\alpha_{m}$, written as $\sigma[\alpha_{1},\ldots,\alpha_{m}]$, is the unique permutation obtained by replacing each entry $\sigma(i)$ by an interval that is order isomorphic to $\alpha_{i}$, where the relative ordering of the intervals corresponds to the ordering of the entries of $\sigma$.
Conversely, a \emph{block-decomposition} or \emph{deflation} of a permutation $\pi$ is any expression of $\pi$ written as an inflation $\pi=\sigma[\alpha_{1},\ldots,\alpha_{m}]$.
For example $521634$ can be decomposed as $3142[1,21,1,12]$. In \Cref{fig:block} we have highlighted the blocks of the permutation $521634$, we can observe that the blocks are placed order isomorphic to $3142$.

A permutation $\pi$ is \emph{plus-decomposable} if it has a block-decomposition of the form $\pi = 12 [\alpha_1,\alpha_2]$ for non-empty permutations $\alpha_1$ and $\alpha_2$ (\Cref{lst:pdecomp}).
For example $213654$ is a plus-decomposable permutation, but $546123$ or $521634$ (which is also not simple) are not.
\Cref{fig:plus} shows the blocks of $213654$, plus-decomposable permutations will always be in this layout (elements in the bottom left quadrant, and elements in the top right quadrant).

\begin{lstlisting}[ language=essence,caption=Essence code which represents plus-decomposable permutations,label=lst:pdecomp]
given length : int
find perm : sequence (size length, injective) of int(1..length)

$ perm is plus-decomposable
such that
    exists sep : int(1..length-1) .
$ There is a split where there is an interval on the "left" where the maximal point is below the lowest point in the interval on the "right".
        max([ perm(i) | i : int(1..sep) ]) < min([ perm(i) | i : int(sep+1..length) ])
\end{lstlisting}

A permutation $\pi$ is \emph{minus-decomposable} if it has a block-decomposition of the form $\pi = 21[\alpha_1,\alpha_2]$ for non-empty permutations $\alpha_1$ and $\alpha_2$ (\Cref{lst:mdecomp}).
For example $546123$ is a minus-decomposable permutation.
But $213654$ (which is plus-decomposable) or $236145$ (which is not simple) are not.
\Cref{fig:minus} contains the minus-decomposable permutation $546123$.
All minus-decomposable permutations will have a similar layout, with elements in the top left quadrant, followed by elements in the bottom right quadrant.
\begin{lstlisting}[ language=essence,caption=Essence code which represents minus-decomposable permutations,label=lst:mdecomp]
given length : int
find perm : sequence (size length, injective) of int(1..length)

$ perm is minus-decomposable
such that
    exists sep : int(1..length-1) .
$ There is a split where there is an interval on the "left" where the minimal point is above the highest point in the interval on the "right".
        min([ perm(i) | i : int(1..sep) ]) > max([ perm(i) | i : int(sep+1..length) ])
\end{lstlisting}

A permutation $\pi\in S_n$ is \emph{block-wise simple} if and only if it has no interval which can be decomposed into $12[\alpha_1,\alpha_2]$ or $21[\alpha_1,\alpha_2]$.
This property was introduced in \cite{bagno2023blockwise} alongside an equivalent recursive recursive definition.
We represent this property as a constraint model in \Cref{lst:blocksimple}.
For example $2413[3142, 1, 1, 1] = 4253716$ is block-wise simple (\Cref{fig:bsimp}), but $24513$ is not as the interval $45$ can be decomposed into $12[1,1]$ (\Cref{fig:nbsimp}).

\begin{lstlisting}[ language=essence,caption=Essence code which represents block-simple permutations,label=lst:blocksimple]
given length : int
find perm : sequence (size length, injective) of int(1..length)

such that
$ It does not decompose into 12 (i.e. there is a point in the interval that is above the smallest point on the right)
    [ and([ max([ perm(i) | i : int(start..middle) ]) > min([ perm(i) | i : int(middle+1..end) ]) 
$ It does not decompose into 21 (i.e. there is a point in the interval that is below the highest point on the right)
          , min([ perm(i) | i : int(start..middle) ]) < max([ perm(i) | i : int(middle+1..end) ]) 
          ])
$ The below is the setup of the intervals, and the indices
    | start, middle, end : int(1..length)
    , start <= middle
    , middle < end
    , letting minSE be min([ perm(i) | i : int(start..end) ])
    , letting maxSE be max([ perm(i) | i : int(start..end) ])
    , maxSE - minSE = end - start
    ]
\end{lstlisting}

A fixed point of a permutation $\pi$ is an integer $i$ such that $\pi(i)=i$.
A \emph{derangement} is a permutation  with no fixed points (\Cref{lst:derange}).
$4312$ is a derangement whereas $1234$ is not.
As shown in \Cref{fig:deran} none of the elements of the permutation $4312$ are on the red diagonal (which represents $\pi(i)=i$).
\begin{lstlisting}[ language=essence,caption=Essence code which represents derangements,label=lst:derange]
given length : int
find perm : sequence (size length, injective) of int(1..length)

such that
    forAll i : int(1..length) .
$ None of the indices are fixed points.
        perm(i) != i
\end{lstlisting}

Similarly, a \emph{nonderangement} is a permutation with at least one fixed point (\Cref{lst:nond}).
$2431$ is a non-derangement whereas $4321$ is not.
The plot of $2431$ in \Cref{fig:nond} shows $\pi(3)=3$ lies on the diagonal.

\begin{lstlisting}[ language=essence,caption=Essence code which represents nonderangements,label=lst:nond]
given length : int
find perm : sequence (size length, injective) of int(1..length)

such that
$ At least one of the points is fixed.
    exists i : int(1..length) .
        perm(i) = i
\end{lstlisting}

A permutation $\pi\in S_n$ is called an \emph{involution} if $\pi=\pi^{-1}$, or equivalently if $\forall i,j.\  \pi(i)=j \iff \pi(j)=i$ (see \Cref{lst:inv} for the model).
$1243$ is an involution but $2431$ is not.
\Cref{fig:invo} contains the plot of the involution $1243$.

\begin{lstlisting}[ language=essence,caption=Essence code which represents involutions,label=lst:inv]
given length : int
find perm : sequence (size length, injective) of int(1..length)

$ perm is an involution
such that
    forAll i : int(1..length) .
$ perm * perm = identity permutation (increasing permutation)
        perm(perm(i)) = i
\end{lstlisting}

A permutation $\pi\in S_n$ is said to have \emph{parity} if the values at odd indexes are odd and the values in even indexes are even, i.e. $\pi(i) = i \mod 2,\ \forall i \in \{1,\ldots,n\}$ (\Cref{lst:parity}).
For example the permutation $3412$ has parity, whereas $2413$ as in the latter for example the value $2$ (even) is at index $1$ which is odd.
\Cref{fig:parity} and \Cref{fig:nonp} show the plots of the two example permutations.

\begin{lstlisting}[ language=essence,caption=Essence code which represents permutations with parity,label=lst:parity]
given length : int
find perm : sequence (size length, injective) of int(1..length)

such that
    forAll i : int(1..length) .
$ Check for odd/even by doing modulo 2 arithmetic.
        (perm(i) % 2) = i % 2
\end{lstlisting}

\subsection{Permutation Statistics}
Currently we support 5 different permutation statistics, where we count the occurrences of some property of the elements of the permutation.
The constraint solvers will output the statistics alongside the solution permutations.
The statistics we currently consider are ascents, descents, inversions, excedances and the Major index (as listed in \Cref{tab:contri}).

These statistics can be collected as part of finding the solutions, or constrained.
These constraints on the statistics can be arbitrary arithmetic statements, for example permutations with 5 descents, or the number of descents and ascents added together is a multiple of three.

An \emph{ascent} in a permutation $\sigma$ is an index $i$ such that $\sigma(i) < \sigma(i + 1)$.
Similarly, a \emph{descent} is an \index $i$ such that $\sigma(i) > \sigma(i + 1)$.
A pair of indices $(i,j)$ in a permutation $\sigma$ such that $i < j$ and $\sigma(i) > \sigma(j)$ is called an \emph{inversion}.
An \emph{excedance} is an index where $\sigma(i) > i$.
So for a permutation $\sigma$ the statistics are
\begin{itemize}
    \item the number of \emph{inversions} inv$(\sigma) = |\{(i,j): i < j \text{ and } \sigma(i) > \sigma(j)\}|$
    \item the number of \emph{descents} des$(\sigma) = |\{i : \sigma(i) > \sigma(i + 1)\}|$
    \item the number of \emph{ascents} asc$(\sigma) = |\{i : \sigma(i) < \sigma(i + 1)\}|$
    \item the number of \emph{excedances} exc$(\sigma) = |\{i : \sigma(i) > i\}|$.
\item the \emph{Major index}, which is the sum of the positions of the descents maj$(\sigma) = \sum_{\sigma(i)>\sigma(i+1)} i$.
\end{itemize}
In \Cref{lst:stats} we have the code for all five statistics in one.

As an example, in the permutation $\sigma=7164523$, inv$(\sigma)=14$, des$(\sigma)=3$, asc$(\sigma)=3$, exc$(\sigma)=2$ and maj$(\sigma)=9$.

\begin{lstlisting}[ language=essence,caption=Essence code which represents all statistics,label=lst:stats]
given length : int
find perm : sequence (size length, injective) of int(1..length)

$ the inversion count
find inversionCount : int(0..length**2) $ We need to each statistic it a reasonable upper bound
such that
    inversionCount =
$ For each occurence of a larger value is found before a smaller one, we add a 1 to a list and we then sum them all up.
        sum([ 1 | i,j : int(1..length), i < j, perm(i) > perm(j)])

$ the descent count
find descentCount : int(0..length)
such that
    descentCount =
$ For each occurence of a descent, we add a 1 to a list and we then sum them all up.
        sum([ 1 | i : int(1..length-1), perm(i) > perm(i+1)])

$ the ascent count
find ascentCount : int(0..length)
such that
    ascentCount =
$ For each occurence of an ascent, we add a 1 to a list and we then sum them all up.
        sum([ 1 | i : int(1..length-1), perm(i) < perm(i+1)])

$ the excedance count
find excedanceCount : int(0..length)
such that
    excedanceCount =
$ For each occurence of an excedance, we add a 1 to a list and we then sum them all up.
        sum([ 1 | i : int(1..length), perm(i) > i])

$ the majorIndex
find majorIndex : int(0..length**2)
such that
    majorIndex =
$ We follow the sum by the definition of Major Index.
        sum([ i | i : int(1..length-1), perm(i) > perm(i+1) ])
\end{lstlisting}

As mentioned before, the statistics can be turned into properties of a permutation, e.g. rather than checking how many descents a permutation has, we ask for a permutation with a given number of descents.
In \Cref{lst:nrinv} we see the statistic of inversions of a permutation turned into a constraint model that looks for permutations with a given number of inversions.

\begin{lstlisting}[ language=essence,caption=Essence code which represents permutations with a given number of inversions,label=lst:nrinv]
given length : int
given nInversions : int
find perm : sequence (size length, injective) of int(1..length)

such that
$ The given number of inversions is equal to the number of inversions found in the permutation   
    nInversions =
        sum([ 1 | i,j : int(1..length), i < j, perm(i) > perm(j)])
\end{lstlisting}

\section{Composability in Action}\label{sec:composability}

We demonstrate the composability of the models using an illustrative (albeit hypothetical) example scenarios and one example which extends the enumeration of solutions to a conjecture in \cite{CLAESSON20121680}.
The latter model gives rise to 2 new conjectures.

\subsection{Hypothetical Example}
\label{sec:figex}
Suppose we are a permutation pattern researcher seeking insights into the set of permutations that classically avoid the permutation $1324$.
It is worth noting that the growth function for this set of permutations has yet to be discovered; see OEIS \cite[A061552]{oeis}.
A mathematician working on this problem will adopt some assumptions, enumerate permutations under these assumptions and refine these assumptions by studying the enumeration.
Given the rapid growth of the numbers, we examine prevalent patterns and properties within our permutations, either contained or avoided, and aim to filter them out.
To illustrate how the CP approach easily cuts down the number of solutions without having to go through generating permutations and then filtering them we have split this example into 4 steps, in each step we add more constraints (patterns and/or properties).
We then show how many solution permutations are found at each step in \Cref{tab:count}, with \textbf{step 4} containing all properties (which is the full model for our example problem).
As the permutation pattern researcher we would only work with \textbf{step 4}, rather than illustrative which contain fewer constraints.

In the context of exploring permutations that traditionally avoid the permutation $1324$, let us assume we have finished exploring permutations that contain the following mesh pattern: \\ $(213,\{(0,0),(0,1),(1,0),(1,0)\})$, so from now on we want to focus on permutations which avoid this pattern (\textbf{step 1}, Av$(\{1324,(213,\{(0,0),(0,1),(1,0),(1,0)\})\})$).
Consider that we decide to look at permutations which contain both the classic pattern $132$ and the mesh pattern $(123,\{(1,2),(2,1),(1,3),(3,1)\})$ (\textbf{step 2}, Co$(\{132,(123,\{(1,2),(2,1),(1,3),(3,1)\})\})$).

We instantiate a constraint programming model that combines the four patterns above.
We can also add properties, for example: minus-decomposable and involution (\textbf{step 3}, minus-decomposable; \textbf{step 4}, involution).

Existing approaches allow a limited form of compositionality: they support the enumeration of permutations that avoid a set of (mesh) patterns simultaneously, as every pattern type can be turned into a mesh pattern.
So the traditional approach would be to enumerate the set of permutations which avoid the two patterns $1324$ and $(213,\{(0,0),(0,1),(1,0),(1,0)\})$.
We would then need to iterate over this enumeration and filter permutations to identify those containing the patterns $132$ and $(123,\{(1,2),(2,1),(1,3),(3,1)\})$.

As an example, for length $9$ there are $9!=362880$ permutations of which, 4862 avoid the two patterns (\textbf{step 1}).
2841 of these permutations then contain the other two patterns (\textbf{step 2}).
1865 of these are minus-decomposable (\textbf{step 3}) and finally only 19 are also involutions (\textbf{step 4}).
Using a composed constraint programming model for this problem we are able to directly enumerate the $19$ permutations of interest without going through several levels of generate-and-test.
We conduct the same experiment on permutation lengths ranging from 5 to 16.
\Cref{tab:count} shows the number of permutations enumerated at each step.
This table demonstrates that this technique is more efficient than generate-and-test based methods, as the numbers between the columns/steps decrease rapidly.
For some permutation lengths (14--16), generate-and-test is not feasible because earlier steps cannot be fully enumerated within the 1-hour time limit we gave to Minion (a CP solver) \cite{minion}, but the combination of all properties can be enumerated.

In addition to only enumerating permutations of interest, we can ask for permutation statistics to be listed together with the enumeration as well.
For example we can request the number of descents and the number of inversions for each the permutations in the final set within the same constraint programming model.

\begin{table}
\begin{center}
    \begin{tabular}{r|cccr}
    \multicolumn{1}{c|}{\multirow{2}[3]{*}{Length}} & \multicolumn{4}{c}{Number of permutations found} \bigstrut[b]\\
         & Step 1 & Step 2 & Step 3 & Step 4 \bigstrut\\
    \hline
    5     & \multicolumn{1}{r}{42} & \multicolumn{1}{r}{8} & \multicolumn{1}{r}{2} & 0 \bigstrut[t]\\
    6     & \multicolumn{1}{r}{132} & \multicolumn{1}{r}{41} & \multicolumn{1}{r}{19} & 1 \\
    7     & \multicolumn{1}{r}{429} & \multicolumn{1}{r}{180} & \multicolumn{1}{r}{102} & 2 \\
    8     & \multicolumn{1}{r}{1,430} & \multicolumn{1}{r}{730} & \multicolumn{1}{r}{455} & 9 \\
    9     & \multicolumn{1}{r}{4,862} & \multicolumn{1}{r}{2,841} & \multicolumn{1}{r}{1,865} & 19 \\
    10    & \multicolumn{1}{r}{16,796} & \multicolumn{1}{r}{10,815} & \multicolumn{1}{r}{7,321} & 53 \\
    11    & \multicolumn{1}{r}{58,786} & \multicolumn{1}{r}{40,700} & \multicolumn{1}{r}{28,096} & 106 \\
    12    & \multicolumn{1}{r}{208,012} & \multicolumn{1}{r}{152,325} & \multicolumn{1}{r}{106,555} & 255 \\
    13    & \multicolumn{1}{r}{742,900} & \multicolumn{1}{r}{568,883} & \multicolumn{1}{r}{401,729} & 493 \\
    14    & -     & -     & -     & 1,118 \\
    15    & -     & -     & -     & 2,120 \\
    16    & -     & -     & -     & 4,664 \\
    \end{tabular}
\end{center}
    \caption{Running the 4 steps on various permutation lengths\label{tab:count}. Dash indicates timeout after 1-hour of Minion search}
\end{table}%

\subsection{1324-avoiding permutations with a fixed number of inversions}
\label{sec:avinv}
Let us now look at the impact the constraint programming approach has on a combination of property type and property that was investigated by \cite{CLAESSON20121680}.
We will focus on the conjecture 13 stated in \cite{CLAESSON20121680}.
This involves combining the 1324 classical pattern avoidance with a fixed number of inversions.

\begin{conj}
\cite{CLAESSON20121680}

For all nonnegative integers $n$ and $k$, we have $S_n^k(1324) \leq S_{n+1}^{k}(1324)$ (where $S_n^k(1324)$ is the set of permutations of length $n$ with $k$ inversions and which avoid $1324$).
\end{conj}

For example, the set $S_5^9(1324)= \{45321, 53421, 54231, 54312\}$ is the set of permutations of length 5 with each with 9 inversions and avoiding 1324 classically.

In the paper \cite{CLAESSON20121680} present results of an exhaustive enumeration algorithm, which found results of all inversions for permutations up to length 15.
Unfortunately, the authors do not disclose how the algorithm worked, but we hypothesise that it might have been a generate and test approach. Which first generates all 1324 avoiding permutations and then filtered for the number of inversions.

As discussed before the constraint programming approach avoids this and thus we have been able to expand the enumeration further, and give additional insights into the conjecture. Further, there was no need for a specialised algorithm or coding to test this conjecture, we only had to combine our previously created model pieces for classical avoidance and counting inversions. While we are unable to fully prove the conjecture, our results demonstrate further support for the result.

Looking at \Cref{tab:avinv} (this is a pattern that can be observed in \cite{CLAESSON20121680} as well) there is a point at which the number of permutations per inversion count does not change.
In other words, there is a point at which it does not matter what the length ($n$) of the permutation is the number of permutations which avoid 1324 and have a given number of inversion is the same as $n+1$.
To confirm this we have expanded the calculations to the permutations of length 16 and all possible number of inversions.
Further, we are able to strongly suggest at which point this enumeration does not change, up to 20 inversions.

We enumerated the permutations for this problem for permutation lengths from 1 to 25 and for the number of inversions ranging from 0 to 20.
We used a large compute server with 256 cores and 1TB of memory for this task. 
Each enumeration is allowed to use up to 250 cores on this machine. 
The longest running instance (length 23 and number of inversions 20) took just over 5 days in this parallel computing setting, which is the equivalent of running the same enumeration on a single core for 3.5 years.
This parallel setting highlights a significant advantage of employing CP: the feasibility of leveraging high-performance, parallel CP solvers, such as Minion, which can take advantage of  the widely accessible multi-core machine without requiring any special development.

\Cref{tab:avinv} only contains some of these results, the full enumeration of up to length 16 (over all inversions) and up to length 23 (up to 20 inversions) can be found in our repository~\cite{ozgur_akgun_2023_10215929}.

\begin{conj}
For a fixed integer $k$, if $n > k + 2$, then $|S_n^{k}(1324)| = |S_{k+2}^k(1324)|$.
\end{conj}

Further we have evidence that there is an identifiable sequence which the stabilising points create.

\begin{conj}
Let $n$ be a given length of permutations, then $S_n^{k}(1324)$ for $k\in\{0,\ldots,n-2\}$ will be the first $n-1$ entries of \cite[A000712]{oeis}.
\end{conj}

\section{Conclusion}\label{sec:conclusion}

In this paper, we have presented a new approach to enumerating permutations under multiple conditions by leveraging constraint programming.
This approach allows for a declarative definition of permutations properties and pattern avoidance/containment conditions.
Conditions and properties can be arbitrarily composed.
We have demonstrated the versatility of this approach by modelling the pattern avoidance/containment by applying it to 2 examples, one of which demonstrates how the constraint programming approach avoids having to generate a large number of permutations before filtering for the ones based on the additional properties/constraints.

In the second example this paper demonstrated the utility of the library of constraint models by extending the computational results from \cite{CLAESSON20121680}, to find more evidence towards their conjecture as well as identifying 2 new conjectures. This allowed us to leverage the parallel support of modern CP solvers to solve in 5 days what would have taken 3.5 years on a single CPU core.

This work contributes to the growing body of research on permutation enumeration.
The constraint programming based approach complements existing computational tools, offering an alternative method that allows for greater flexibility when solving non-standard permutation enumeration problems.

The application of constraint programming to this important field in mathematics empowers mathematicians with a new flexible tool for investigating complex permutation enumeration problems.
We expect our approach to inspire further research in this area and potentially lead to new mathematical discoveries.

\newgeometry{bottom=10mm}
\begin{landscape}
\begin{table}
\begin{center}
    \begin{tabular}{r|rrrrrrrrrrrrrrrrrrrrrr}
    \multicolumn{1}{c|}{\multirow{2}[3]{*}{Length}} & \multicolumn{21}{c}{Number of permutations found with a fixed number of inversions} \bigstrut[b]\\
 & 0& 1& 2& 3& 4& 5& 6& 7& 8& 9& 10& 11& 12& 13& 14& 15& 16& 17& 18& 19& 20  \bigstrut\\
    \hline

1 & 1 &   &   &    &    &    &    &     &     &     &     &     &      &      &      &      &      &      &       &       &\\
2 & \cellcolor[gray]{0.7} 1 & 1 &   &    &    &    &    &     &     &     &     &     &      &      &      &      &      &      &       &       &\\
3 & 1 & \cellcolor[gray]{0.7}2 & 2 & 1  &    &    &    &     &     &     &     &     &      &      &      &      &      &      &       &       &\\
4 & 1 & 2 &\cellcolor[gray]{0.7} 5 & 6  & 5  & 3  & 1  &     &     &     &     &     &      &      &      &      &      &      &       &       &\\
5 & 1 & 2 & 5 & \cellcolor[gray]{0.7}10 & 16 & 20 & 20 & 15  & 9   & 4   & 1   &     &      &      &      &      &      &      &       &       &\\
6 & 1 & 2 & 5 & 10 & \cellcolor[gray]{0.7}20 & 32 & 51 & 67  & 79  & 80  & 68  & 49  & 29   & 14   & 5    & 1    &      &      &       &       &\\
7 & 1 & 2 & 5 & 10 & 20 & \cellcolor[gray]{0.7}36 & 61 & 96  & 148 & 208 & 268 & 321 & 351  & 347  & 308  & 241  & 165  & 98   & 49    & 20    & 6     \\
8 & 1 & 2 & 5 & 10 & 20 & 36 & \cellcolor[gray]{0.7}65 & 106 & 171 & 262 & 397 & 568 & 784  & 1019 & 1264 & 1478 & 1628 & 1681 & 1619  & 1441  & 1173  \\
9 & 1 & 2 & 5 & 10 & 20 & 36 & 65 & \cellcolor[gray]{0.7}110 & 181 & 286 & 443 & 664 & 985  & 1416 & 1988 & 2715 & 3589 & 4579 & 5631  & 6654  & 7559  \\
10& 1 & 2 & 5 & 10 & 20 & 36 & 65 & 110 & \cellcolor[gray]{0.7}185 & 296 & 467 & 714 & 1077 & 1582 & 2305 & 3284 & 4617 & 6374 & 8665  & 11521 & 15012 \\
11& 1 & 2 & 5 & 10 & 20 & 36 & 65 & 110 & 185 & \cellcolor[gray]{0.7}300 & 477 & 738 & 1127 & 1682 & 2477 & 3584 & 5134 & 7240 & 10100 & 13915 & 18976 \\
12& 1 & 2 & 5 & 10 & 20 & 36 & 65 & 110 & 185 & 300 & \cellcolor[gray]{0.7}481 & 748 & 1151 & 1732 & 2577 & 3768 & 5450 & 7766 & 10976 & 15312 & 21171 \\
13& 1 & 2 & 5 & 10 & 20 & 36 & 65 & 110 & 185 & 300 & 481 & \cellcolor[gray]{0.7}752 & 1161 & 1756 & 2627 & 3868 & 5634 & 8098 & 11526 & 16216 & 22632 \\
14& 1 & 2 & 5 & 10 & 20 & 36 & 65 & 110 & 185 & 300 & 481 & 752 & \cellcolor[gray]{0.7}1165 & 1766 & 2651 & 3918 & 5734 & 8282 & 11858 & 16786 & 23568 \\
15& 1 & 2 & 5 & 10 & 20 & 36 & 65 & 110 & 185 & 300 & 481 & 752 & 1165 & \cellcolor[gray]{0.7}1770 & 2661 & 3942 & 5784 & 8382 & 12042 & 17118 & 24138 \\
16& 1 & 2 & 5 & 10 & 20 & 36 & 65 & 110 & 185 & 300 & 481 & 752 & 1165 & 1770 & \cellcolor[gray]{0.7}2665 & 3952 & 5808 & 8432 & 12142 & 17302 & 24470 \\
17& 1 & 2 & 5 & 10 & 20 & 36 & 65 & 110 & 185 & 300 & 481 & 752 & 1165 & 1770 & 2665 & \cellcolor[gray]{0.7}3956 & 5818 & 8456 & 12192 & 17402 & 24654 \\
18& 1 & 2 & 5 & 10 & 20 & 36 & 65 & 110 & 185 & 300 & 481 & 752 & 1165 & 1770 & 2665 & 3956 & \cellcolor[gray]{0.7}5822 & 8466 & 12216 & 17452 & 24754 \\
19& 1 & 2 & 5 & 10 & 20 & 36 & 65 & 110 & 185 & 300 & 481 & 752 & 1165 & 1770 & 2665 & 3956 & 5822 & \cellcolor[gray]{0.7}8470 & 12226 & 17476 & 24804 \\
20& 1 & 2 & 5 & 10 & 20 & 36 & 65 & 110 & 185 & 300 & 481 & 752 & 1165 & 1770 & 2665 & 3956 & 5822 & 8470 & \cellcolor[gray]{0.7}12230 & 17486 & 24828 \\
21& 1 & 2 & 5 & 10 & 20 & 36 & 65 & 110 & 185 & 300 & 481 & 752 & 1165 & 1770 & 2665 & 3956 & 5822 & 8470 & 12230 & \cellcolor[gray]{0.7}17490 & 24838 \\
22& 1 & 2 & 5 & 10 & 20 & 36 & 65 & 110 & 185 & 300 & 481 & 752 & 1165 & 1770 & 2665 & 3956 & 5822 & 8470 & 12230 & 17490 & \cellcolor[gray]{0.7}24842 \\
23& 1 & 2 & 5 & 10 & 20 & 36 & 65 & 110 & 185 & 300 & 481 & 752 & 1165 & 1770 & 2665 & 3956 & 5822 & 8470 & 12230 & 17490 & 24842 \\ \hline
A000712 & 1 & 2 & 5 & 10 & 20 & 36 & 65 & 110 & 185 & 300 & 481 & 752 & 1165 & 1770 & 2665 & 3956 & 5822 & 8470 & 12230 & 17490 & 24842
    \end{tabular}%
\end{center}

    \caption{The enumeration of permutations avoiding 1324 classically and containing a fixed number of inversions\label{tab:avinv}. The rows are the length of the permutation, the columns the number of inversions.
    The full results can be found in the supplementary repository~\cite{ozgur_akgun_2023_10215929}. The last row shows the first 21 elements of the sequence A000712 \cite[A000712]{oeis}.}
\end{table}%
\end{landscape}
\restoregeometry

\nocite{*}
\bibliographystyle{abbrvnat}
\bibliography{bibliography}

\begin{thebibliography}{27}
\providecommand{\natexlab}[1]{#1}
\providecommand{\url}[1]{\texttt{#1}}
\expandafter\ifx\csname urlstyle\endcsname\relax
  \providecommand{\doi}[1]{doi: #1}\else
  \providecommand{\doi}{doi: \begingroup \urlstyle{rm}\Url}\fi

\bibitem[Akg{\"u}n et~al.(2022)Akg{\"u}n, Frisch, Gent, Jefferson, Miguel, and
  Nightingale]{akgun2022conjure}
{\"O}.~Akg{\"u}n, A.~M. Frisch, I.~P. Gent, C.~Jefferson, I.~Miguel, and
  P.~Nightingale.
\newblock {Conjure: Automatic generation of constraint models from problem
  specifications}.
\newblock \emph{Artificial Intelligence}, 310:\penalty0 103751, 2022.

\bibitem[Akg{\"{u}}n et~al.(2022)Akg{\"{u}}n, Mereb, and
  Vendramin]{akgun2022enumeration}
{\"{O}}.~Akg{\"{u}}n, M.~Mereb, and L.~Vendramin.
\newblock {Enumeration of set-theoretic solutions to the Yang-Baxter equation}.
\newblock \emph{Math. Comput.}, 91\penalty0 (335):\penalty0 1469--1481, 2022.
\newblock URL \url{https://doi.org/10.1090/mcom/3696}.

\bibitem[Akg{\"{u}}n et~al.(2023)Akg{\"{u}}n, Hoffmann, and
  Jefferson]{ozgur_akgun_2023_10215929}
{\"{O}}.~Akg{\"{u}}n, R.~Hoffmann, and C.~Jefferson.
\newblock {stacs-cp/composable-permutation-patterns}, Nov. 2023.
\newblock URL \url{https://doi.org/10.5281/zenodo.10215929}.

\bibitem[Albert(2012)]{albert2012permlab}
M.~Albert.
\newblock {PermLab: Software for permutation patterns}, 2012.
\newblock URL \url{https://github.com/mchllbrt/PermCode}.

\bibitem[Albert et~al.(2012)Albert, Linton, and Hoffmann]{PatternClass}
M.~Albert, S.~Linton, and R.~Hoffmann.
\newblock {PatternClass--Permutation Pattern Classes}, 2012.
\newblock URL \url{https://gap-packages.github.io/PatternClass/}.

\bibitem[Ardal et~al.(2021)Ardal, Magnusson, Émile Nadeau, Kristinsson,
  Gudmundsson, Bean, Ulfarsson, Eliasson, Tannock, Bjarnason, Pantone, and
  Arnarson]{Permuta}
R.~P. Ardal, T.~K. Magnusson, Émile Nadeau, B.~J. Kristinsson, B.~A.
  Gudmundsson, C.~Bean, H.~Ulfarsson, J.~S. Eliasson, M.~Tannock, A.~B.
  Bjarnason, J.~Pantone, and A.~B. Arnarson.
\newblock {Permuta}, Apr. 2021.
\newblock URL \url{https://doi.org/10.5281/zenodo.4725759}.

\bibitem[Atkinson(1999)]{DBLP:journals/dm/Atkinson99}
M.~D. Atkinson.
\newblock {Restricted permutations}.
\newblock \emph{Discret. Math.}, 195\penalty0 (1-3):\penalty0 27--38, 1999.
\newblock URL \url{https://doi.org/10.1016/S0012-365X(98)00162-9}.

\bibitem[Avgustinovich et~al.(2013)Avgustinovich, Kitaev, and
  Valyuzhenich]{boxedmesh}
S.~Avgustinovich, S.~Kitaev, and A.~Valyuzhenich.
\newblock {Avoidance of boxed mesh patterns on permutations}.
\newblock \emph{Discrete Applied Mathematics}, 161\penalty0 (1-2):\penalty0
  43--51, Jan. 2013.
\newblock ISSN 0166-218X.
\newblock URL \url{https://doi.org/10.1016/j.dam.2012.08.015}.

\bibitem[Babson and Steingr{\'\i}msson(2000)]{babson2000generalized}
E.~Babson and E.~Steingr{\'\i}msson.
\newblock {Generalized permutation patterns and a classification of the
  Mahonian statistics}.
\newblock \emph{S{\'e}minaire Lotharingien de Combinatoire [electronic only]},
  44:\penalty0 B44b--18, 2000.

\bibitem[Bagno et~al.(2023)Bagno, Eisenberg, Reches, and
  Sigron]{bagno2023blockwise}
E.~Bagno, E.~Eisenberg, S.~Reches, and M.~Sigron.
\newblock {Blockwise simple permutations}, 2023.

\bibitem[Bousquet-M{\'e}lou et~al.(2010)Bousquet-M{\'e}lou, Claesson, Dukes,
  and Kitaev]{bousquet20102}
M.~Bousquet-M{\'e}lou, A.~Claesson, M.~Dukes, and S.~Kitaev.
\newblock {(2+ 2)-free posets, ascent sequences and pattern avoiding
  permutations}.
\newblock \emph{Journal of Combinatorial Theory, Series A}, 117\penalty0
  (7):\penalty0 884--909, 2010.

\bibitem[Br{\"a}nd{\'e}n and Claesson(2011)]{branden2011mesh}
P.~Br{\"a}nd{\'e}n and A.~Claesson.
\newblock {Mesh Patterns and the Expansion of Permutation Statistics as Sums of
  Permutation Patterns}.
\newblock \emph{The Electronic Journal of Combinatorics}, pages P5--P5, 2011.

\bibitem[Brignall(2010)]{brignall2010survey}
R.~Brignall.
\newblock {A survey of simple permutations}.
\newblock \emph{Permutation patterns}, 376:\penalty0 41--65, 2010.

\bibitem[Chu et~al.(2018)Chu, Stuckey, Schutt, Ehlers, Gange, and
  Francis]{chu2018chuffed}
G.~Chu, P.~J. Stuckey, A.~Schutt, T.~Ehlers, G.~Gange, and K.~Francis.
\newblock {Chuffed, a lazy clause generation solver}, 2018.
\newblock URL \url{https://github.com/chuffed/chuffed}.

\bibitem[Claesson et~al.(2012)Claesson, Jelínek, and
  Steingrímsson]{CLAESSON20121680}
A.~Claesson, V.~Jelínek, and E.~Steingrímsson.
\newblock {Upper bounds for the Stanley–Wilf limit of 1324 and other layered
  patterns}.
\newblock \emph{Journal of Combinatorial Theory, Series A}, 119\penalty0
  (8):\penalty0 1680--1691, 2012.
\newblock ISSN 0097-3165.
\newblock URL
  \url{https://www.sciencedirect.com/science/article/pii/S0097316512000891}.

\bibitem[Distler et~al.(2012)Distler, Jefferson, Kelsey, and
  Kotthoff]{semigroupcount}
A.~Distler, C.~Jefferson, T.~Kelsey, and L.~Kotthoff.
\newblock {The Semigroups of Order 10}.
\newblock In M.~Milano, editor, \emph{Principles and Practice of Constraint
  Programming}, pages 883--899, Berlin, Heidelberg, 2012. Springer Berlin
  Heidelberg.
\newblock ISBN 978-3-642-33558-7.

\bibitem[E{\'e}n and S{\"o}rensson(2003)]{een2003extensible}
N.~E{\'e}n and N.~S{\"o}rensson.
\newblock {An extensible SAT-solver}.
\newblock In \emph{International conference on theory and applications of
  satisfiability testing}, pages 502--518. Springer, 2003.

\bibitem[Frisch et~al.(2008)Frisch, Harvey, Jefferson,
  Mart{\'\i}nez-Hern{\'a}ndez, and Miguel]{frisch2008ssence}
A.~M. Frisch, W.~Harvey, C.~Jefferson, B.~Mart{\'\i}nez-Hern{\'a}ndez, and
  I.~Miguel.
\newblock {Essence: A constraint language for specifying combinatorial
  problems}.
\newblock \emph{Constraints}, 13\penalty0 (3):\penalty0 268--306, 2008.

\bibitem[GAP()]{GAP4}
GAP.
\newblock \emph{{GAP -- Groups, Algorithms, and Programming, Version 4.12.2}}.
\newblock The GAP~Group, 2022.
\newblock URL \url{https://www.gap-system.org}.

\bibitem[Gent et~al.(2006)Gent, Jefferson, and Miguel]{minion}
I.~P. Gent, C.~Jefferson, and I.~Miguel.
\newblock {Minion: A Fast Scalable Constraint Solver}.
\newblock In \emph{ECAI}, pages 98--102, 2006.

\bibitem[Knuth(1968)]{knuth}
D.~E. Knuth.
\newblock \emph{{The Art of Computer Programming, Volume {I:} Fundamental
  Algorithms}}.
\newblock Addison-Wesley, 1968.

\bibitem[Magnusson and Ulfarsson(2012)]{magnusson2012algorithms}
H.~Magnusson and H.~Ulfarsson.
\newblock {Algorithms for discovering and proving theorems about permutation
  patterns}.
\newblock \emph{arXiv preprint arXiv:1211.7110}, 2012.

\bibitem[Marcus and Tardos(2004)]{marcus2004excluded}
A.~Marcus and G.~Tardos.
\newblock {Excluded permutation matrices and the Stanley--Wilf conjecture}.
\newblock \emph{Journal of Combinatorial Theory, Series A}, 107\penalty0
  (1):\penalty0 153--160, 2004.

\bibitem[Nadeau et~al.(2021)Nadeau, Bean, Ulfarsson, Eliasson, and
  Pantone]{CombExp}
{\'{E}}.~Nadeau, C.~Bean, H.~Ulfarsson, J.~S. Eliasson, and J.~Pantone.
\newblock {Combinatorial Specification Searcher
  (Permuta{T}riangle/comb\_spec\_searcher): Version 4.0.0}, June 2021.
\newblock URL \url{https://doi.org/10.5281/zenodo.4946832}.

\bibitem[Nightingale et~al.(2017)Nightingale, Akg{\"u}n, Gent, Jefferson,
  Miguel, and Spracklen]{nightingale2017automatically}
P.~Nightingale, {\"O}.~Akg{\"u}n, I.~P. Gent, C.~Jefferson, I.~Miguel, and
  P.~Spracklen.
\newblock {Automatically improving constraint models in Savile Row}.
\newblock \emph{Artificial Intelligence}, 251:\penalty0 35--61, 2017.

\bibitem[{OEIS Foundation Inc.}(2023)]{oeis}
{OEIS Foundation Inc.}
\newblock {The {O}n-{L}ine {E}ncyclopedia of {I}nteger {S}equences}, 2023.
\newblock Published electronically at \url{http://oeis.org}.

\bibitem[Simion and Schmidt(1985)]{simion1985restricted}
R.~Simion and F.~W. Schmidt.
\newblock {Restricted permutations}.
\newblock \emph{European Journal of Combinatorics}, 6\penalty0 (4):\penalty0
  383--406, 1985.

\end{thebibliography}
\label{sec:biblio}

\end{document}